\documentclass[twocolumn,aps,prb,superscriptaddress,floatfix,noeprint]{revtex4-1}

\usepackage{graphicx,tabularx}
\usepackage{amsmath,amsfonts,amssymb}
\usepackage{bm,dsfont}
\usepackage[colorlinks=true, citecolor=blue, linkcolor=red, urlcolor=blue]{hyperref}
\usepackage[all]{hypcap}
\usepackage{xcolor}
\usepackage[caption=false]{subfig}

\graphicspath{{figures/}}

\DeclareMathOperator{\sgn}{sgn}

\newcolumntype{C}{>{\centering\arraybackslash}X}

\begin{document}

\title{Effective Floquet model for minimally twisted bilayer graphene}
\author{Christophe De Beule}
\affiliation{Institute for Mathematical Physics, TU Braunschweig, 38106 Braunschweig, Germany}
\affiliation{Department of Physics and Materials Science, University of Luxembourg, L-1511 Luxembourg, Luxembourg}
\author{Fernando Dominguez}
\affiliation{Institute for Mathematical Physics, TU Braunschweig, 38106 Braunschweig, Germany}
\author{Patrik Recher}
\affiliation{Institute for Mathematical Physics, TU Braunschweig, 38106 Braunschweig, Germany}
\affiliation{Laboratory for Emerging Nanometrology, 38106 Braunschweig, Germany}
\date{\today}

\begin{abstract}
We construct an effective Floquet lattice model for the triangular network that emerges in interlayer-biased minimally twisted bilayer graphene and which supports two chiral channels per link for a given valley and spin. We introduce the Floquet scheme with the one-channel triangular network and subsequently extend it to the two-channel case. From the bulk topological index (winding number) and finite system calculations, we find that both cases host anomalous Floquet insulators (AFIs) with a different gap-opening mechanism. In the one-channel network, either time-reversal or in-plane inversion symmetry has to be broken to open a gap. In contrast, in the two-channel network, interchannel coupling can open a gap without breaking these symmetries yielding a valley AFI with counterpropagating edge states. This phase is topologically trivial with respect to the total winding number but robust in the absence of intervalley scattering. Finally, we demonstrate the applicability of the Floquet model with magnetotransport calculations.
\end{abstract}

\maketitle

\section{Introduction}

Stacking two graphene sheets with a relative twist produces a moir\'e pattern that drastically alters the electronic structure \cite{LopesDosSantos2007,SuarezMorell2010,Bistritzer2010,Li2010}. At the magic angle ($\theta \sim 1^\circ$) the low-energy bands are almost flat, such that many-body effects dominate giving rise to superconductivity and strongly-correlated phases \cite{Kim2017a,Cao2018a,Cao2018,Yankowitz2019,Sharpe2019,Kerelsky2019,Choi2019,Cao2019}.  When the twist angle is reduced far below the magic angle, the system exhibits markedly different behavior. In this limit, the moir\'e pattern is reconstructed into a triangular tiling of AB and BA stacking domains \cite{Nam2017,Yoo2019,Walet2019} whose vertices are given by AA regions acting as topological defects \cite{Alden2013}. At such tiny twist angles ($\theta \sim 0.1^\circ$), the system is referred to as minimally twisted bilayer graphene (mTBG). Furthermore, when layer-inversion symmetry is broken in mTBG, for example, by a perpendicular electric field, a local gap is opened in the Bernal regions with a different topological character for AB and BA stacking. In particular, the change in valley Chern number across an AB/BA domain wall is quantized to $\pm 2$. Each domain wall therefore supports two chiral modes for a given valley and spin that counterpropagate for different valleys \cite{Martin2008,Zhang2013,Vaezi2013,Ju2015,Yin2016}.  When the Fermi level lies in the gap, the low-energy physics is derived solely from a triangular network of valley Hall states \cite{San-Jose2013,Efimkin2018,Ramires2018,Huang2018,Sunku2018,Rickhaus2018,Xu2019}. Here, the AA regions remain metallic and correspond to the scattering nodes of the network. These nodes are connected by links given by the AB/BA domain walls, where each link hosts two chiral channels for a given valley and spin that scatter at the nodes, see Fig.\ \ref{fig:intro}(a). One thus obtains an oriented triangular scattering network for each valley, where the orientation is opposite for opposite valleys. This regime in mTBG can thus be modeled by a Chalker-Coddington-like network model \cite{Chalker1988} for each valley separately \cite{Efimkin2018}. Recently, it was demonstrated that the network in mTBG gives rise to a triplet of one-dimensional (1D) chiral zigzag modes \cite{Fleischmann2020,Tsim2020}. Importantly, the chiral zigzag modes require interchannel scattering at the nodes, such that one necessarily needs to consider a two-channel model to capture the network physics in mTBG \cite{DeBeule2020}. 

Oriented scattering networks were initially introduced as models for the percolation transition between two quantum Hall plateaus \cite{Chalker1988,Kramer2005,Mkhitaryan2009,Lee1994}. Recently, it was demonstrated that oriented networks can also be mapped to periodically driven (Floquet) systems \cite{Pasek2014,Delplace2017,Delplace2020}. Such Floquet systems can host anomalous insulating phases that support topological boundary modes even though the bulk topology is trivial \cite{Kitagawa2010,Liang2013,Rudner2014,Mukherjee2020}. Hence, it stands to reason that scattering networks can also host anomalous edge modes \cite{Delplace2020,Chou2020}. The connection between these two viewpoints has been addressed recently in Ref.\ \onlinecite{Potter2020}. In this paper, we construct an effective Floquet lattice model for the oriented triangular scattering network that emerges in mTBG under interlayer bias. Our motivation is threefold. Firstly, the topological properties of the gapped phases in the two-channel model have not yet been addressed. Moreover, a Hamiltonian description is preferable here because the bulk-edge correspondence is ill defined for scattering networks \cite{Delplace2020}. Secondly, an effective tight-binding model for the network in mTBG is highly desirable. As the number of atoms in a moir\'e cell is of the order of $10^4 \, (\theta^\circ)^{-2}$, it has similar advantages over atomistic methods as network models. Additionally, it is straightforward to implement using standard codes. Finally, there is the prospect of reproducing the equilibrium physics of the network in a driven system with photonic crystals \cite{Raghu2008,Lu2014,Ozawa2019,Zhong2020} or cold atoms in optical lattices \cite{Lewenstein2007,Bloch2008,Bakr2009,Wintersperger2020}. 

The paper is organized as follows: In Sec.\ \ref{sec:1channel}, we introduce our approach with the one-channel triangular network. We start by mapping a single scattering node of the network to the time evolution of a three-level system, i.e., a trimer. We then show how the network dynamics can be reproduced by a triangular lattice of trimers and construct the Floquet Hamiltonian. We find three distinct phases: a metal, a trivial insulator, and an anomalous Floquet insulator (AFI) where a gap can only be opened by either breaking time-reversal or in-plane inversion symmetry. We then proceed to the two-channel network in Sec.\ \ref{sec:2channel}, which is the general case realized in mTBG. Besides metallic phases that support chiral zigzag modes, we find that the two-channel network hosts AFIs within a single valley even in the presence of time reversal and inversion symmetry where now the gap is opened via interchannel coupling. Taking into account both valleys, we show that the two-channel network exhibits a valley AFI with counterpropagating edge modes characterized by valley winding numbers. Finally, we showcase the applicability of the effective Floquet model by performing transport calculations in Sec.\ \ref{sec:transport}, and we present our conclusions in Section \ref{sec:end}.

\section{One-channel network} \label{sec:1channel}

\begin{figure}
\centering
\includegraphics[width=\linewidth]{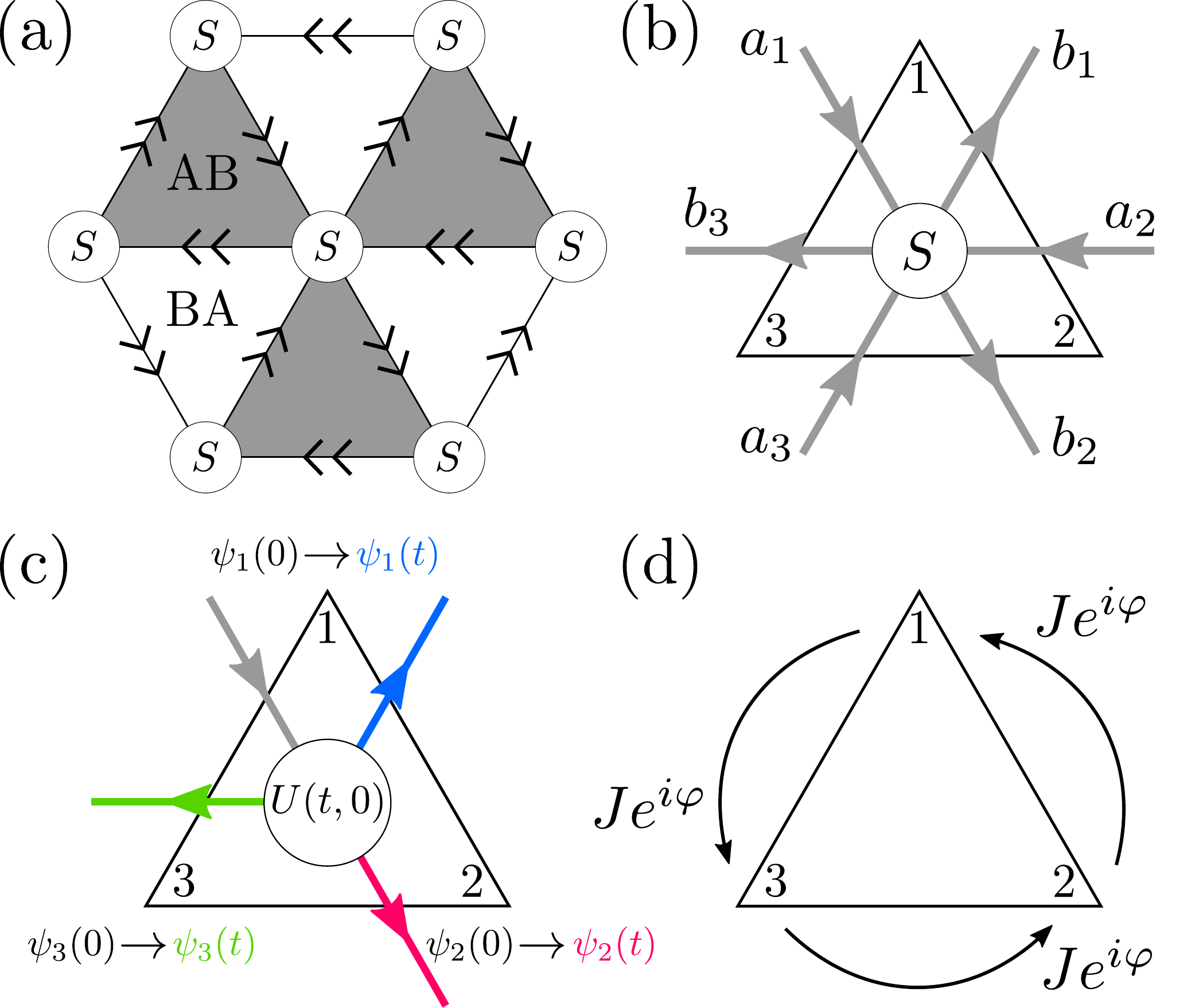}
\caption{(a) Oriented triangular network in mTBG with two channels per link for a given valley and spin. (b) Scattering at a triangular node where incoming and outgoing modes are identified with the time evolution of a trimer, whose sites are denoted by $1$, $2$, and $3$. (c) Mapping between the time evolution of trimer amplitudes and the $S$ matrix. (d) Trimer hoppings in the counterclockwise direction [Eq.\ \eqref{eq:ham0}].}
\label{fig:intro}
\end{figure}

We start by considering the simplest case where the two valley Hall states are decoupled. In this case, the scattering network is given by two copies of a single-channel triangular network with $C_3$ symmetry for a given valley and spin [Fig.\ \ref{fig:intro}(a)]. Here, we do not take into account intervalley scattering as the moir\'e pattern in mTBG varies slowly on the interatomic scale. Note also that the orientation of the network is opposite for the two valleys as they are related by time reversal. Moreover, mTBG under interlayer bias has an additional $C_2T$ symmetry that conserves the valley, where $C_2$ corresponds to in-plane inversion with respect to a node, which reverses both the valley and sublattice pseudospin, and $T$ is the time-reversal operator \cite{Zou2018}. This combined symmetry can for example be broken by a magnetic field or an in-plane electric field. The one-channel triangular network with $C_2T$ symmetry was first considered by Efimkin and MacDonald \cite{Efimkin2018}, while the general case with broken $C_2T$ has recently been considered in Ref.\ \onlinecite{Chou2020}, both within the framework of network models. Here, we show explicitly how these scattering networks can be mapped to Floquet tight-binding models with discrete time steps.

\subsection{Scattering node}

To illustrate the basic idea, we first consider a single scattering node with three incoming and three outgoing modes, as shown in Fig.\ \ref{fig:intro}(b). The outgoing modes $b=(b_1,b_2,b_3)^t$ are related to the incoming modes $a=(a_1,a_2,a_3)^t$ by the $S$ matrix, $b=Sa$. If $C_3$ symmetry is preserved, the $S$ matrix can be written as
\begin{equation} \label{eq:S1}
S=
\begin{pmatrix}
s_l & s_r & s_f \\
s_f & s_l & s_r \\
s_r & s_f & s_l
\end{pmatrix},
\end{equation}
where $|s_f| ^2=P_f$, $|s_l|^2=P_l$, and $|s_r| ^2=P_r$ are, respectively, the probability for forward scattering, and left and right deflections. Current conservation at the node is expressed by $S^\dag S = \mathds 1_3$ such that up to a global phase, the $S$ matrix only depends on two real parameters with $P_f + P_r + P_l=1$. For the special case where $C_2T$ symmetry is preserved, this is reduced to a single parameter. For example, we can take $s_r=s_l=\sqrt{P_d}$ and $s_f=e^{-i\alpha}\sqrt{P_f}$ with $\cos \alpha = -\sqrt{P_d/4P_f}$, such that $1/9 \leq P_f \leq 1$.

\begin{figure}
\centering
\includegraphics[width=.9\linewidth]{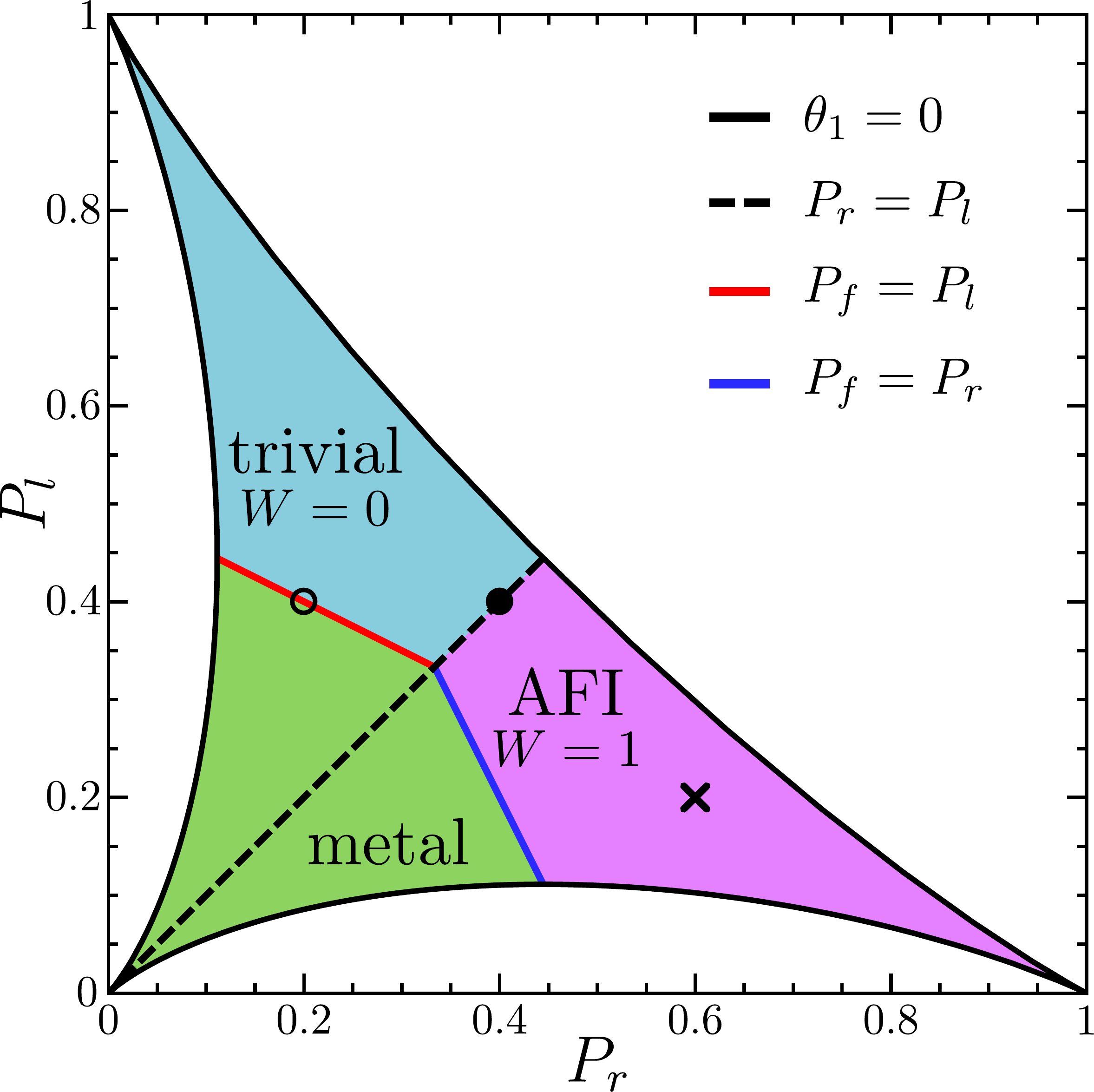}
\caption{Phase diagram of the oriented one-channel triangular network corresponding to the driven trimer lattice shown in Fig.\ \ref{fig:trimer}(a) where different colors correspond to different phases as indicated. The boundary of the allowed $(P_r,P_l)$ values corresponds to $\theta_1=0$ in \eqref{eq:Pl} and the dashed line gives the case with $C_2T$ symmetry ($P_r=P_l$). If the network orientation is reversed, the trivial and AFI phase are interchanged.}
\label{fig:phase1}
\end{figure}

We now demonstrate how the scattering problem can be mapped to the time-evolution of a trimer, i.e., a three-level system corresponding to a particle hopping between the vertices of an equilateral triangle with amplitude $\psi=(\psi_1,  \psi_2, \psi_3)^t$, as shown in Fig.\ \ref{fig:intro}(b). For a time-independent Hamiltonian $H_0$, the amplitudes evolve in time as $\psi(t) = U(t,0) \psi(0)$ where $U(t,0) = e^{-iH_0t/\hbar}$. If the scattering process takes place over a time $t$, we can make the following identification \cite{Delplace2020}
\begin{equation}
b=Sa \quad \leftrightarrow \quad \psi(t) = U(t,0) \psi(0),
\end{equation}
which is illustrated in Fig.\ \ref{fig:intro}(c) for an initial state $\psi(0)=(1,0,0)^t$ corresponding to an incoming mode at the upper vertex. After a time $t$, we have $\psi(t) =  (s_l,s_f,s_r)^t$ which is interpreted as the scattering amplitudes for a left deflection, forward scattering, or a right deflection, respectively. The general $S$ matrix with $C_3$ symmetry is obtained by taking the following trimer Hamiltonian,
\begin{equation} \label{eq:ham0}
H_0 = J \begin{pmatrix} 0 & e^{i\varphi} & e^{-i\varphi} \\ e^{-i\varphi} & 0 & e^{i\varphi} \\ e^{i\varphi} & e^{-i\varphi} & 0 \end{pmatrix} ,
\end{equation}
with $J\geq0$ and where $3\varphi$ is the flux through the triangular plaquette [Fig.\ \ref{fig:intro}(d)]. Up to an overall energy shift, this is the most general $C_3$-symmetric trimer Hamiltonian. The spectrum and eigenstates of \eqref{eq:ham0} are given by
\begin{equation}
E_n = 2 J \cos \left( \varphi + 2\pi n/3 \right), \,\,\,\,
\psi_n = \frac{1}{\sqrt{3}} \left( \eta^n, \eta^{-n} , 1 \right)^t,
\end{equation}
with $\eta=e^{i2\pi /3}$ and $n=0,1,2$. Next, we identify the time-evolution operator $U(t)=e^{-iH_0t/\hbar}$ with the $S$ matrix given by Eq.\ \eqref{eq:S1}. This gives $s_l = g(\theta_1,\theta_2)$,  $s_r = g(\theta_1,\theta_2-2\pi/3)$, and $s_f = g(\theta_1,\theta_2+2\pi/3)$ with
\begin{equation} \label{eq:g}
g(\theta_1,\theta_2) = \frac{1}{3} \left( e^{-i\theta_1} + 2 e^{i\theta_1} \cos \theta_2 \right),
\end{equation}
and
\begin{equation} \label{eq:theta}
\theta_1 = \frac{3}{2} \, J t \cos \varphi / \hbar, \qquad \theta_2 = \sqrt{3} \, J t \sin \varphi /\hbar,
\end{equation}
where we left out an overall phase factor in Eq.\ \eqref{eq:g} that only depends on $\theta_1$. We thus have
\begin{equation}  \label{eq:Pl}
P_l =  \frac{1}{9} \left[ 3 + 4 \cos(2\theta_1) \cos \theta_2 + 2 \cos(2 \theta_2) \right],
\end{equation}
and where $P_f$ and $P_r$ are obtained from \eqref{eq:Pl} by letting $\theta_2 \rightarrow \theta_2 \pm 2\pi/3$, respectively. Note that the scattering probabilities are functions of $Jt$ and $\varphi$. This is illustrated in Fig.\ \ref{fig:phase1}, where we plot $(P_r,P_l)$ for $\theta_2 \in [0,2\pi]$ and $\theta_1=0$, which bounds the allowed region of the scattering parameters \cite{Chou2020}. Regions with different colors in the figure correspond to different phases, which is explained in Section \ref{sec:topo}. Note that the map $(\theta_1,\theta_2) \rightarrow (P_r,P_l)$ is onto, but not one-to-one. Apart from an overall phase, the sign of $\theta_1$ has to be specified to uniquely determine the $S$ matrix.

In the special case where $C_2T$ symmetry is conserved, we further require that $s_r =s_l$ which is the case for $\theta_2 = (n + 1/3) \pi$ with $n$ an integer. For example, for $\theta_2=\pi/3$, the amplitudes become $s_f = - \tfrac{1}{3} \cos \theta_1 - i \sin\theta_1$ and  $s_r=s_l= \tfrac{2}{3} \cos(\theta_1)$. The resulting $S$ matrix is unitary equivalent to the one of Efimkin and MacDonald \cite{Efimkin2018}.
\begin{figure*}
\centering
\includegraphics[width=.9\linewidth]{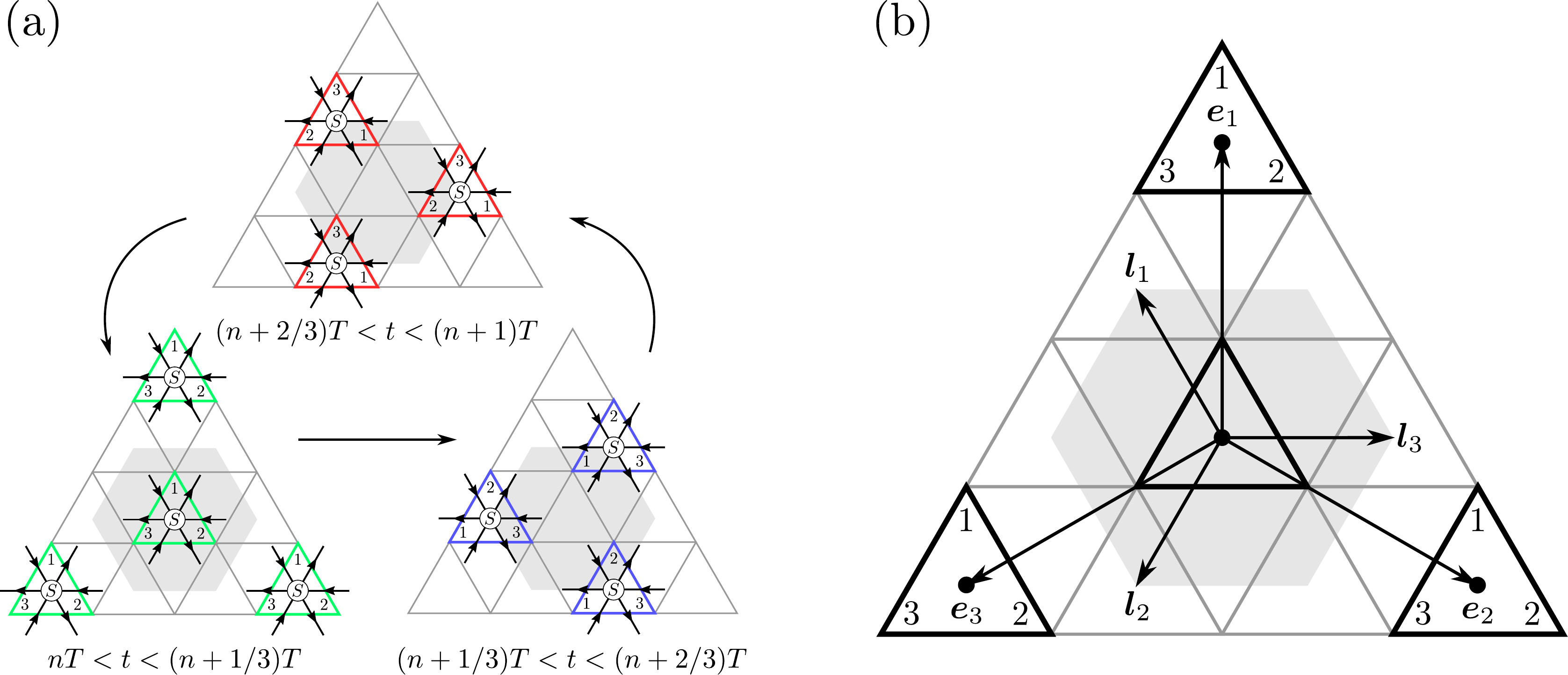} 
\caption{(a) Periodic sequence in which the trimers on the lattice are turned on and off, where the couplings between different sites of a trimer are indicated by the thick colored lines. This generates the dynamics of the oriented scattering network that is superimposed. (b) Triangular lattice of trimers whose vertices are labeled by $1$, $2$, and $3$. Here, the vertices of different cells are connected by thin gray lines and the unit cell is shown as the gray area.}
\label{fig:trimer}
\end{figure*}

\subsection{Scattering network}

Having established the mapping for a single node, we turn to the scattering network. To this end, we follow Ref.\ \onlinecite{Delplace2020} and decompose the network into three disjoint sets of scattering nodes. The nodes are then identified with three sets of decoupled trimers that form a triangular lattice, as illustrated in the different panels of Fig.\ \ref{fig:trimer}(a). Note that a scattering process on one set takes as input the output of another set. For example, in the figure the incoming modes of green trimers (bottom-left panel) correspond to the outgoing modes of blue trimers (bottom-right panel). If we only turn on the trimer couplings in one set for a given time, the time evolution generates a local scattering process. Next, we do the same for a different set, which takes the output of the first set, thereby transporting the amplitudes through the lattice, followed by a scattering process at the new nodes. Repeating this process in the sequence shown in Fig.\ \ref{fig:trimer}(a) reproduces the network dynamics. Note that the orientation of the network is fixed by the specific sequence in which the couplings are switched on and off, which naturally breaks time-reversal symmetry. Thus, by exchanging the second and third step, the orientation of the network is reversed.

The piecewise time-dependent Hamiltonian can be written as
\begin{equation} \label{eq:ham1}
H(\bm k,t) = \begin{cases}
H_1(\bm k), \quad & \quad 0 < t < T/3, \\ 
H_2(\bm k), \quad & \quad T/3 < t < 2T/3, \\
H_3(\bm k), \quad & \quad 2T/3 < t < T,
\end{cases}
\end{equation}
with $H(\bm k,t+T)=H(\bm k,t)$ and where $\bm k$ is the Bloch momentum of the trimer lattice. We now explicitly construct the Hamiltonians $H_1$, $H_2$, and $H_3$ for each step and show that the network dynamics is reproduced.

Consider a triangular lattice of decoupled trimers, as shown in Fig.\ \ref{fig:trimer}(b). Each unit cell contains three sites that constitute a trimer centered at $\bm r_{mn} = m \bm e_1 + n \bm e_2$ ($m,n\in \mathbb Z$) with
\begin{equation}
\bm e_1=\sqrt{3} \, l (0,1), \qquad \bm e_2 = \sqrt{3} \, l (\sqrt{3}/2,-1/2).
\end{equation}
where the lattice constant is given by $\sqrt{3} \, l$ with $l$ the link length of the triangular network. As we discussed above, the network dynamics are obtained by a three-step process that is repeated periodically. Because of $C_3$ symmetry, the coupling and duration of each step has to be equal. In the first step, which takes place between times $t=0$ and $t=T/3$, we only turn on green trimers, as illustrated in the bottom-left panel of Fig.\ \ref{fig:trimer}(a). Here, there is no coupling between different cells, such that the Hamiltonian in Bloch form $H_1=H_0$, where $H_0$ is given in Eq.\ \eqref{eq:ham0}, and
\begin{equation}
U_1 = U(T/3,0) = S, 
\end{equation}
with $S=e^{-iH_0T/3\hbar}$. In the scattering picture, the trimer amplitudes at $t=0$ correspond to incoming modes of scattering nodes located at $\bm r_{mn}$. Outgoing modes then correspond to the trimer amplitudes at $t=T/3$. In the second step, we turn off the trimer coupling within each cell and turn on the coupling between vertices of different cells in such a way that we obtain the blue trimers shown in the bottom-right panel of Fig.\ \ref{fig:trimer}(a), which gives 
\begin{equation}
\label{eq:H2}
H_2(\bm k) = 
\begin{pmatrix} 0 & z e^{i\bm k \cdot \bm e_1} & z^*e^{-i\bm k \cdot \bm e_3} \\
z^* e^{-i\bm k \cdot \bm e_1} & 0 & z e^{i\bm k \cdot \bm e_2} \\
z e^{i\bm k \cdot \bm e_3} & z^* e^{-i\bm k \cdot \bm e_2} & 0 \end{pmatrix},
\end{equation}
with $z=Je^{i\varphi}$ and $\bm e_3 = -(\bm e_1+\bm e_2)$. The time-evolution operator for this step becomes
\begin{equation} \label{eq:U2}
U_2 = U(2T/3,T/3) = T_3 T_2 S T_1,
\end{equation}
where
\begin{align}
T_1 & = \textrm{diag} \left( e^{i\bm k \cdot \bm l_2}, e^{i\bm k \cdot \bm l_1}, e^{i\bm k \cdot \bm l_3} \right), \label{eq:T1} \\
T_2 & = \textrm{diag} \left( e^{i\bm k \cdot \bm l_3}, e^{i\bm k \cdot \bm l_2}, e^{i\bm k \cdot \bm l_1} \right), \\
T_3 & = \textrm{diag} \left( e^{i\bm k \cdot \bm l_1}, e^{i\bm k \cdot \bm l_3}, e^{i\bm k \cdot \bm l_2} \right),
\end{align}
with $\bm l_{1,2} = l(-1/2,\pm\sqrt{3}/2)$ and $\bm l_3=-(\bm l_1+\bm l_2)$ primitive vectors of the network [Fig.\ \ref{fig:intro}(a)]. This can be interpreted in terms of the network as follows. Outgoing modes of nodes located at $\bm r_{mn}$, i.e., trimer amplitudes at $t=T/3$, first propagate to the next node via the translation operator $T_1$. Hence, they can be thought of as the incoming modes of nodes located at $\bm r_{mn} - \bm l_j$ ($j=1,2,3$) which then scatter to outgoing modes by $S$, followed again by propagation. The final step takes place during $2T/3<t<T$, giving rise to the red trimers shown in the top panel of Fig.\ \ref{fig:trimer}(a), such that
\begin{equation}
\label{eq:H3}
H _3(\bm k) =
\begin{pmatrix} 0 & ze^{-i\bm k \cdot \bm e_2} & z^*e^{i\bm k \cdot \bm e_1} \\
z^*e^{i\bm k \cdot \bm e_2} & 0 & ze^{-i\bm k \cdot \bm e_3} \\
ze^{-i\bm k \cdot \bm e_1} & z^*e^{i\bm k \cdot \bm e_3} & 0 \end{pmatrix},
\end{equation}
with time-evolution operator
\begin{equation}
U_3 = U(T,2T/3) = T_3 S T_3^\dag,
\end{equation}
which has a similar interpretation as $U_2$. In the end, outgoing modes of nodes located at $\bm r_{mn} + \bm l_j$ ($j=1,2,3$) are propagated by $T_3$, becoming incoming modes of trimers centered at $\bm r_{mn}$. At $t=T$, we therefore end up back where we started at $t=0$. The time-evolution operator over one period (Floquet operator) becomes
\begin{equation} \label{eq:UF}
U_F (\bm k) = \mathcal T e^{-\frac{i}{\hbar}\int_0^T dt \, H(\bm k,t)} = T_3 S T_2 S T_1 S,
\end{equation}
where $\mathcal T$ denotes time ordering.

\begin{figure}
\centering
\includegraphics[width=\linewidth]{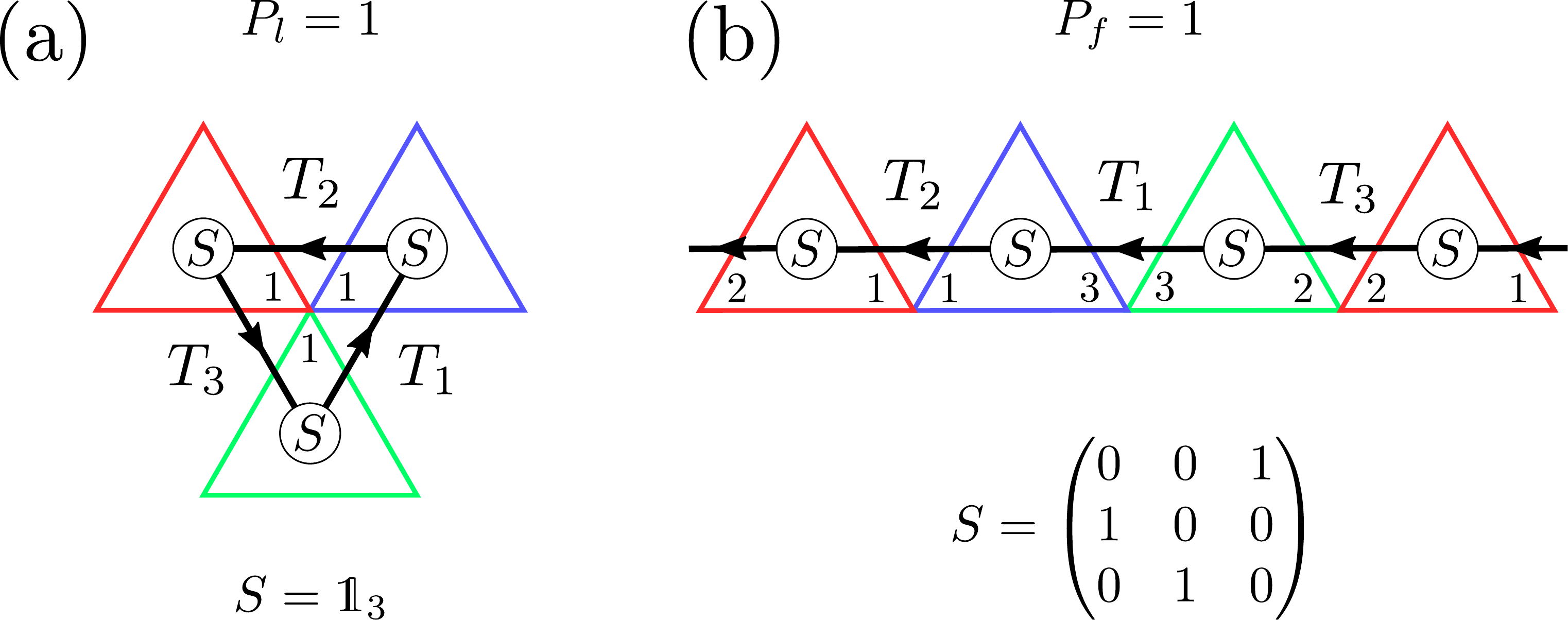}
\caption{Correspondence between the driven trimer lattice and the triangular network. (a) For $P_l=1$ (e.g.\ $\theta_1=\theta_2 = 0$), the network is localized leading to flatbands. (b) For $P_f=1$ (e.g.\ $\theta_1=0$ and $\theta_2 = -2\pi/3$), there are three sets of chiral modes along the $-\bm l_j$ directions (only the $-\bm l_3$ mode is shown).}
\label{fig:network}
\end{figure}

To demonstrate the correspondence of the trimer lattice with the triangular oriented scattering network, we show the case where $P_l=1$ (e.g.\ $\theta_1=\theta_2 = 0$) or $P_f=1$ (e.g.\ $\theta_1=0$ and $\theta_2 = -2\pi/3$) in Fig.\ \ref{fig:network} (a) and (b), respectively. When $P_l=1$ and $P_r=P_f=0$, the network modes perform closed orbits giving an insulator with flatbands. In the opposite limit, we have $P_f=1$ and $P_r=P_l=0$, such that the network is a metal consisting of three sets of 1D chiral modes that propagate along the $-\bm l_j$ directions ($j=1,2,3$).
\begin{figure}
\centering
\includegraphics[width=\linewidth]{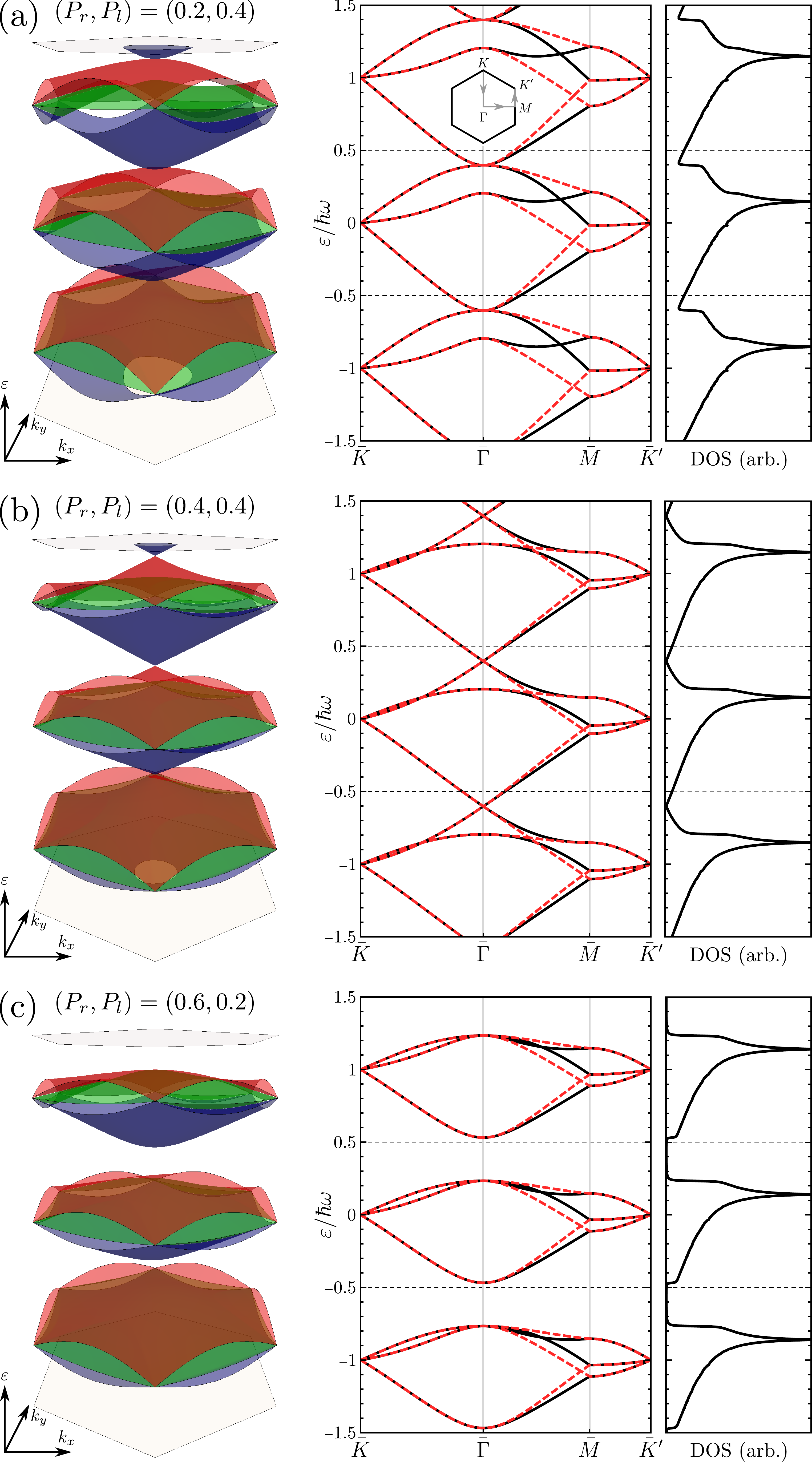}
\caption{Floquet bands in the BZ of the trimer lattice (left panels) and along high-symmetry lines (middle panels), together with the DOS (right panels) for $\gamma/\hbar\omega=0.005$. Here, (a), (b), and (c) are indicated by the circle, dot, and cross in Fig.\ \ref{fig:phase1}, respectively. Solid (dashed) lines correspond to the (opposite) network orientation (i.e., valley index) as shown in Fig.\ \ref{fig:intro}(a).}
\label{fig:bands1}
\end{figure}

\subsection{Quasienergy spectrum}

According to the Floquet theorem, the wave equation $i\hbar\partial_t \Psi_{\bm k}(t)=H(\bm k,t) \Psi_{\bm k}(t)$ with a time-periodic Hamiltonian $H(\bm k,t+T)=H(\bm k,t)$ is solved by
\begin{equation}
\Psi_{\bm k}(t) = e^{-i\varepsilon_{\bm k}t/\hbar} \sum_m e^{im\omega t} \psi_{\bm k m},
\end{equation}
where $\varepsilon_{\bm k}$ is the quasienergy and $\omega=2\pi/T$. The wave equation then gives
\begin{equation}
\sum_{m'} \mathcal H^{mm'} \psi_{\bm km'} = \varepsilon_{
\bm k} \psi_{\bm km},
\end{equation}
where $\mathcal H^{mm'}$ is the Floquet Hamiltonian,
\begin{equation}
\mathcal H^{mm'} = m \hbar \omega \delta_{mm'} \mathds 1_3 + \frac{1}{T} \int_0^T dt \, e^{-i(m-m') \omega t} H(t),
\end{equation}
where we suppressed the momentum index. When $H(t)$ is given by \eqref{eq:ham1}, we find for $m=m'$,
\begin{equation}
\mathcal H^{mm} = m \hbar \omega \mathds 1_3 + \frac{1}{3} \sum_{j=1}^3 H_j,
\end{equation}
and for $m\neq m'$,
\begin{equation} \label{eq:HF2}
\mathcal H^{mm'} = \frac{\sin \left( \frac{\pi \Delta m}{3} \right)}{\pi \Delta m} e^{-i\pi \Delta m /3} \sum_{j=1}^3 H_j e^{\frac{2\pi i}{3} (1-j) \Delta m},
\end{equation}
where $\Delta m = m-m'$. Unless specifically stated, we always use eight harmonics ($m=-8,\ldots,8$) in the Floquet Hamiltonian for numerical calculations to ensure convergence. The scattering parameters of the one-channel model can now be written as
\begin{equation}
\theta_1 = \frac{J}{\hbar \omega} \, \pi \cos \varphi, \qquad \theta_2 = \frac{J}{\hbar \omega} \frac{2\pi}{\sqrt{3}} \sin \varphi.
\end{equation}
In terms of the scattering network,
\begin{equation}
\hbar \omega = \frac{2\pi}{3} \frac{\hbar v}{l} \approx 100 \, \theta^\circ \textrm{meV},
\end{equation}
as $T=3l/v$, where $v$ is the velocity of the chiral modes, $l$ is the link length of the network, and $\theta$ is given in degrees as indicated. Here, we also give the numerical value for the case of mTBG where we put $v$ equal to the bulk Fermi velocity of graphene and $l$ to the moir\'e lattice constant. In the remainder of this paper, we assume that the scattering parameters (for both the one- and two-channel model) are approximately constant on this energy scale, such that they are essentially energy independent.

The Floquet quasienergy spectrum is shown in Fig.\ \ref{fig:bands1} for several values of $(P_r,P_l)$ together with the density of states (DOS). Note that there are three bands per energy period since there are three sites per trimer (labeled $1$, $2$, and $3$ in Fig.\ \ref{fig:trimer}). The DOS is calculated numerically with a Lorentzian broadening $\gamma$,
\begin{align}
\textrm{DOS}(\varepsilon) & = \frac{1}{V} \sum_{m,n} \sum_{\bm k} \delta \left( \varepsilon - \varepsilon_{mn}(\bm k) \right) \\
& \rightarrow \frac{1}{V} \sum_{m,n} \sum_{\bm k} \frac{\gamma/\pi}{( \varepsilon_{mn}(\bm k) - \varepsilon )^2 + \gamma^2},
\end{align}
where $m \in \mathbb Z$, $n=0,1,2$ labels the three bands per Floquet period, and the sum runs over the first Brillouin zone. Note that the lattice constant of the trimer lattice is $\sqrt{3} \, l$ . Thus, the Brillouin zone (BZ) of the trimer lattice is reduced by a factor $3$ as compared to the network BZ. For example, when $C_2T$ is conserved, the origin and the two inequivalent corners of the network BZ support Dirac nodes \cite{Efimkin2018} separated in energy by $2\pi\hbar v/3l$. These nodes are folded to the $\bar \Gamma$ point ($\bm k = 0$) of the trimer lattice BZ. At the $\bar \Gamma$ point we have $H_1=H_2=H_3$ such that different harmonics become decoupled [Eq.\ \eqref{eq:HF2}]. Thus, the spectrum at the origin is given by
\begin{equation} \label{eq:eps0}
\varepsilon_{mn}(0) = m \hbar\omega + 2J \cos \left( \varphi + 2\pi n/3 \right).
\end{equation}
\begin{figure}
\centering
\includegraphics[width=\linewidth]{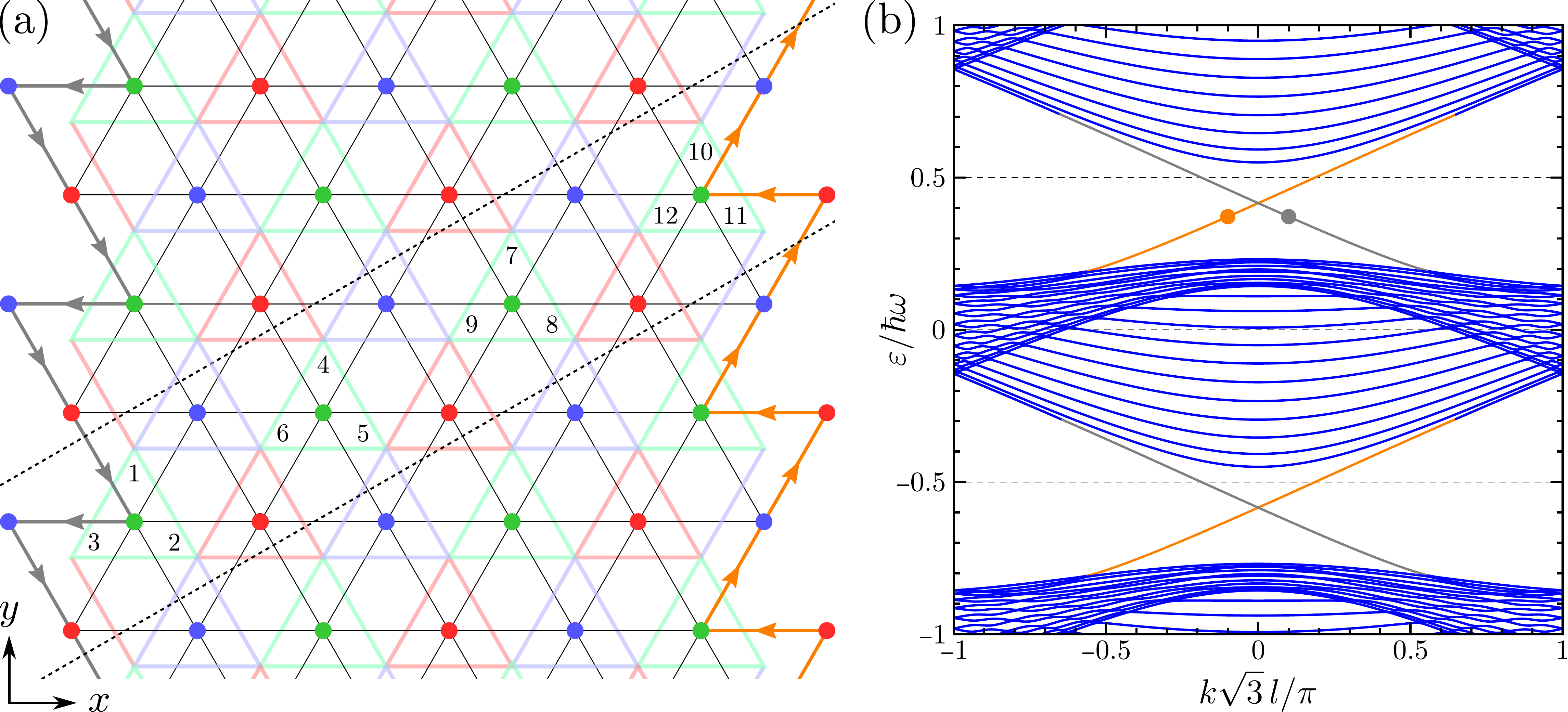}
\caption{(a) Zigzag ribbon of the trimer lattice in the $\bm e_1$ direction with $N=4$ trimers in the ribbon unit cell, indicated by the dashed lines, giving $12$ sites per cell. (b) Spectrum for $N=8$ in the AFI phase with $(P_r,P_l) = (0.6,0.2)$ indicated in Fig.\ \ref{fig:phase1} by the cross, whose bulk bands are shown in Fig.\ \ref{fig:bands1}(c).}
\label{fig:ribbon1}
\end{figure}

Now we can already partly understand the phase diagram shown in Fig.\ \ref{fig:phase1}. By construction, the network for $P_r=P_l=0$ is given by decoupled chiral modes with $\varepsilon_{mj}(\bm k) = \hbar \omega \left( m - 3\bm k \cdot \bm l_j / 2\pi \right)$ ($j=1,2,3$). These modes are coupled when we allow for deflections and at some point a gap opens between different triads of Floquet bands at the $\bar \Gamma$ point, unless $C_2T$ is preserved. We find from \eqref{eq:eps0} that the gap opens when $P_f=P_r$ ($\theta_2 = n\pi$) or $P_f=P_l$ ($\theta_2 = \left(n - 1/3 \right) \pi$) for $P_{l,r}\in[1/9,1/3]$, respectively. The band touching is quadratic in general, as shown in Fig.\ \ref{fig:bands1}(a), while in the special case where $C_2T$ is conserved, it is linear and symmetry protected [Fig.\ \ref{fig:bands1}(b)]. Hence, the phase diagram is given by three distinct regions: one metal and two insulating phases separated by a percolation line ($P_r =P_l$) along which $C_2T$ is conserved \cite{Chou2020}. 

\subsection{Anomalous Floquet phase} \label{sec:topo}

For network models, the bulk-edge correspondence is ill defined because one can always engineer the boundary of the network such that it hosts a chiral mode, even if the bulk is trivial \cite{Delplace2020}. For example, one can surround a bulk network consisting of localized loops with a disconnected large loop which does not alter bulk properties. In contrast, the bulk-edge correspondence of Floquet insulators is well defined \cite{Rudner2014}.
\begin{figure*}
\centering
\includegraphics[width=.9\linewidth]{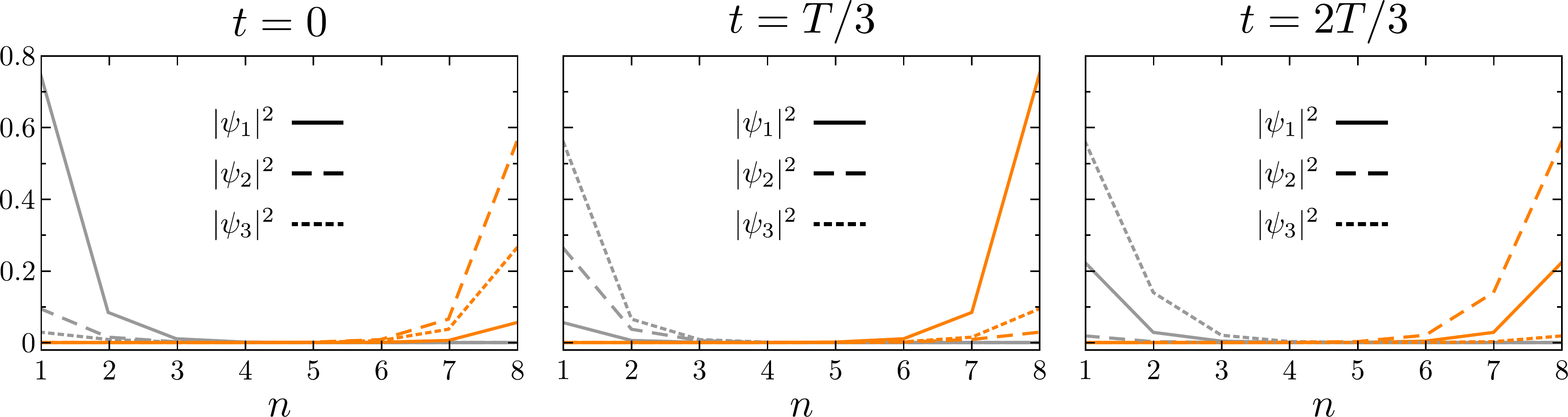}
\caption{Trimer densities of the edge states marked by the gray (orange) dot in Fig.\ \ref{fig:ribbon1}(b) at different times during one driving period, where the edge state on the left (right) edge corresponds to the gray (orange) curves and $n=1,\ldots N$ labels the trimers in the unit cell from left to right, as shown in Fig.\ \ref{fig:ribbon1}(a).}
\label{fig:density1}
\end{figure*}
To establish the topology of the insulating phases, we calculated the Floquet winding number $W$ as well as the spectrum of finite networks. The winding number is a bulk topological invariant characterizing anomalous Floquet phases \cite{Rudner2014}.

In Fig.\ \ref{fig:ribbon1}(a), we show a zigzag ribbon of the trimer lattice, which is finite along the $x$ axis with width $(3N-1)l/2$, where $N$ is the number of trimers in the unit cell, and infinite along the $y$ axis. The Floquet Hamiltonian of the ribbon is given in Appendix \ref{app:ribbon}. The corresponding network is superimposed on the figure, where the nodes and links are shown as the dots and black solid lines, respectively. Note that the edge of the network is automatically determined. Here, the edge has trimers, dimers, and monomers, which correspond to nodes having three, two, or one incoming and outgoing modes, respectively, giving rise to a sawtooth edge. For example, the $S$ matrix corresponding to the edge dimer is given by
\begin{equation}
S_d = \exp \left[ - i \frac{JT}{3} \left( \cos \varphi \, \sigma_x \pm \sin \varphi \, \sigma_y \right) \right],
\end{equation}
for the right and left edge, respectively. Here, we take the same scattering parameters at the edge as in the bulk for simplicity. From Fig.\ \ref{fig:ribbon1}(a), we observe that for $P_r \sim 1$ the edge supports a chiral mode that propagates counterclockwise along the edge, while for $P_l \sim 1$, the edge is localized. This suggests an anomalous Floquet insulator (AFI, $W=1$) and a trivial insulator $(W=0)$, respectively. We confirmed this by calculating the winding number for these two limiting cases (Appendix \ref{app:winding}) which establishes the phase diagram shown in Fig.\ \ref{fig:phase1}.  As expected, the ribbon spectrum in the AFI phase, shown in Fig.\ \ref{fig:ribbon1}(b), features a pair of chiral modes (one for each edge) in each gap. The corresponding probability densities of the three trimer sublattices are shown in Fig.\ \ref{fig:density1} at times $t=0$, $T/3$, and $2T/3$ for the edge states marked in Fig.\ \ref{fig:ribbon1}(b). They correspond to incoming modes of the scattering nodes in the green, blue, and red trimers, respectively. We find that the anomalous edge states propagate mostly along the sawtooth edges. Furthermore, we verified that the edge state in the AFI phase is robust by varying the hopping constants at the edge. 

In mTBG, the network orientation is opposite for valley $K$ and $K'$. If valley $K$ corresponds to the phase diagram shown in Fig.\ \ref{fig:phase1}, then the phase diagram of valley $K'$ is obtained by exchanging the trivial and AFI phase, as well as the sign of $W$. However, the scattering parameters of the two valleys are in general not related: $(P_r, P_l)$ for $K$ differs from $(P_r',P_l')$ for $K'$. The former has a trivial phase for $P_r < P_l$ and an AFI ($W=1$) for $P_r > P_l$, while the latter has a trivial phase for $P_r' > P_l'$ and an AFI ($W=-1$) phase for $P_r' < P_l'$. If either $C_2$ or $T$ is conserved, they are related as follows
\begin{alignat}{2}
C_2: & \qquad s_{r(l)K'} = s_{r(l)K}, \qquad && s_{fK'} = s_{fK}, \\
T: & \qquad s_{r(l)K'} = ( s_{l(r)K} )^t, \qquad && s_{fK'} = ( s_{fK} )^t.
\end{alignat}
Hence, when $T$ is broken and $C_2$ is conserved, if one valley hosts a trivial phase then the other valley always hosts an AFI phase and vice versa. If the valleys are decoupled in the bulk, the system can be thought of as two half cylinders, one for each valley, hosting different topological phases that are glued together, giving rise to anomalous edge states along the seams \cite{Chou2020}. On the other hand if only $C_2$ is broken, then either both valleys host a trivial phase or both host an AFI phase with opposite winding numbers. With respect to the total winding number, these are topologically equivalent. Indeed, in the latter case, intervalley scattering at the boundary will gap out the edge modes. However, there is a weak topological phase characterized by the valley winding numbers with a pair of counterpropagating edge modes that is robust as long as intervalley coupling is absent or small compared to the bulk gap. An overview of the topological phases of the one-channel triangular network is shown in Table \ref{tab:phases}.
\begin{table}
\centering
\begin{ruledtabular}
	\begin{tabular}{c|ccc}
 	\hphantom{aa}$N_c$\hphantom{aa} & $C_2$ & $T$ & $C_2T$ \\[.5mm]
	\hline \\[-3mm]
	$1$ & $(1,0)$ or $(0,-1)$ & $(1,-1) \sim (0,0)$ & n.a.\ \\
	$2$ & -- & -- & $(1,-1) \sim (0,0)$
	\end{tabular}
\end{ruledtabular}
\caption{Anomalous topological phases hosted by the triangular network with $N_c$ channels per valley and spin. Pairs indicate the valley winding numbers when the symmetry displayed in the first row is conserved. Blank entries were not considered in this work. Here, $\sim$ indicates topological equivalence with respect to the total winding number.}
\label{tab:phases}
\end{table}

\section{Two-channel network} \label{sec:2channel}

The Floquet model for the two-channel network is constructed using a similar approach as for the single-channel network. Here, the second channel is introduced by taking trimers with two orbitals per site. In the presence of $C_3$ symmetry about the center, the general Hamiltonian for a trimer with two orbitals can be written as
\begin{equation} \label{eq:ham2}
H_0 = \begin{pmatrix} h_1 & h_{12} \\ h_{12}^\dag & h_2 \end{pmatrix},
\end{equation}
in the basis $\psi = \left( \psi_{11} , \psi_{12}, \psi_{13} , \psi_{21}, \psi_{22} , \psi_{23} \right)^t$, where the first (second) index denotes the orbital (site). We also have, up to an orbital-independent energy shift,
\begin{equation}
h_j = (-1)^j \delta \, \mathds 1_3 + \begin{pmatrix} 0 & z_j & z_j^* \\ z_j^* & 0 & z_j \\ z_j & z_j^* & 0 \end{pmatrix}, \, h_{12} = \begin{pmatrix} z_5 & z_3 & z_4^* \\ z_4^* & z_5 & z_3 \\ z_3 & z_4^* & z_5 \end{pmatrix},
\end{equation}
with $z_j = J_j e^{i\varphi_j}$. Here, $h_1$ and $h_2$ contain the intraorbital hoppings $z_1$ and $z_2$, and on-site energies $\pm \delta$. Interorbital couplings are given by $h_{12}$ with hoppings $z_3$ and $z_4$, and on-site terms $z_5$ (Fig.\ \ref{fig:2coupling} in the Appendix). This gives a total of eleven parameters and the time-evolution operator becomes intractable analytically. We therefore opt for a different approach where we start from the desired $S$ matrix and numerically compute an effective trimer Hamiltonian:
\begin{equation}
\frac{H_0}{\hbar \omega} = \frac{3i}{2\pi} \log S,
\end{equation}
such that $S=e^{-iH_0T/3\hbar}$. The form of $H_0$ depends on the branch cut of the logarithm, but this is unimportant as it results in the same network dynamics. 

In a previous work \cite{DeBeule2020}, we demonstrated that the $S$ matrix of the two-channel oriented triangular network in the presence of $C_3$ and $C_2T$ can be written in the form given by Eq.\ \eqref{eq:S1} with
\begin{align}
s_f & = \begin{pmatrix} e^{i(\phi+\chi)}\sqrt{P_{f1}} & -\sqrt{P_{f2}} \\ -\sqrt{P_{f2}} & -e^{-i(\phi+\chi)}\sqrt{P_{f1}} \end{pmatrix}, \\
s_r & = \begin{pmatrix} e^{i\phi}\sqrt{P_{d1}} & \sqrt{P_{d2}} \\ -\sqrt{P_{d2}} & -e^{-i\phi}\sqrt{P_{d1}} \end{pmatrix},
\end{align}
and $s_l=(s_r)^t$, where $P_{f1}+P_{f2}+2(P_{d1}+P_{d2})=1$. Here, $P_{f1}$ ($P_{f2}$) and $P_{d1}$ ($P_{d2}$) are the intrachannel (interchannel) forward scattering probability and deflection probability, respectively. The relative phase shift between intrachannel deflections of the two channels equals $2\phi+\pi$ and $\cos \chi = \left( P_{d2} - P_{d1} \right)/2\sqrt{P_{f1} P_{d1}}$ with $2\sqrt{P_{f1}P_{d1}} \geq \left| P_{d2} - P_{d1} \right|$ such that $\chi$ is real. Hence, we have four phenomenological scattering parameters in total, which can be chosen as $P_{f1}$, $P_{f2}$, $\phi$, and $P_{d1}-P_{d2}$. Note that this is not the most general $S$ matrix. Indeed, we assumed that the intrachannel scattering probabilities for the two channels are equal, as well as taking equal probabilities for interchannel deflections to the left and right.

We thus have $(b_1,b_1',b_2,b_2',b_3,b_3')^t = S (a_1,a_1',a_2,a_2',a_3,a_3')^t$ where $a$ and $a'$ are the amplitudes of the three incoming modes of the two channels, respectively, and similar for outgoing modes $b$ and $b'$. The components are defined similarly as before, e.g., $a_1$ and $a_1'$ both propagate along the downward diagonal link [Fig.\ \ref{fig:intro}(b)]. For simplicity, we consider $P_d=P_{d1}=P_{d2}$ in the remainder of this work.
\begin{figure}
\centering
\includegraphics[width=\linewidth]{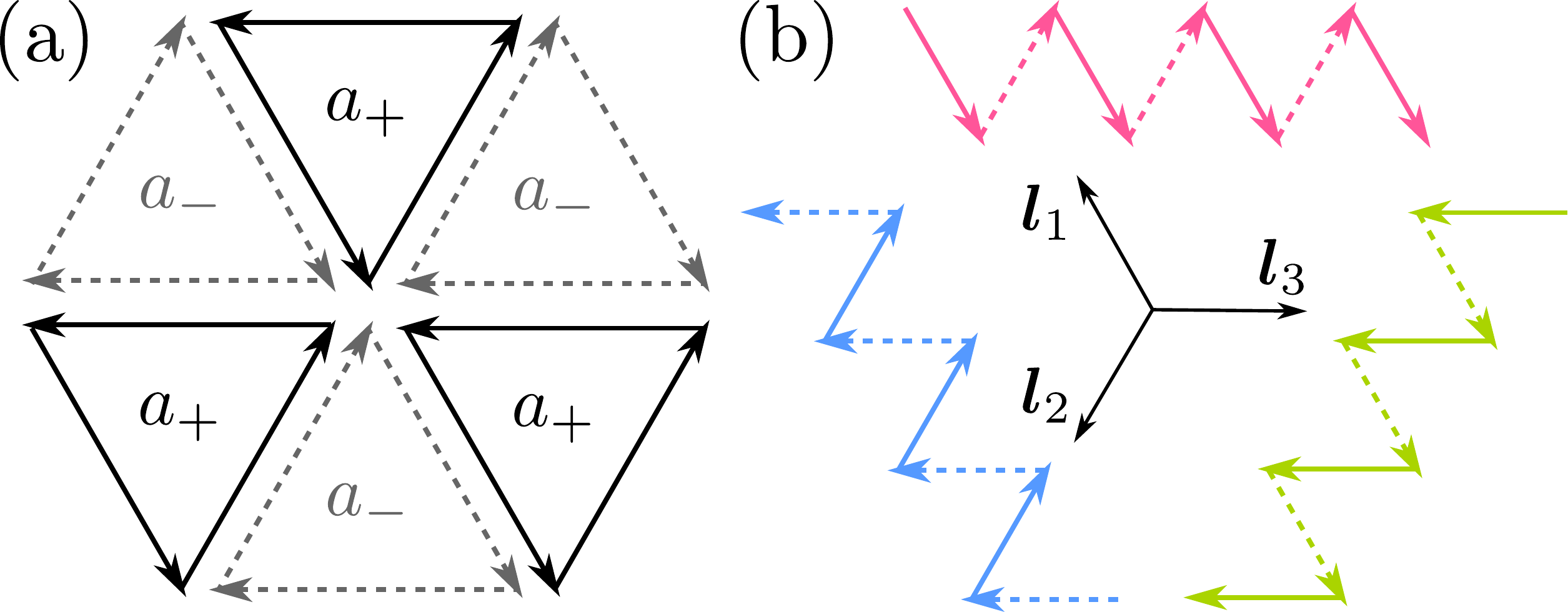}
\caption{(a) Pseudo-Landau levels for $\phi = (n+1/2)\pi$ where solid (dashed) arrows correspond to $a_+$ ($a_-$) superpositions of valley Hall states along the same link. (b) Triplet of 1D chiral zigzag modes for $\phi = n\pi$ that propagate in the $\bm l_j$ directions.}
\label{fig:2channel}
\end{figure}

\subsection{Absence of forward scattering}

We first consider the case without forward scattering ($s_f=0$). This is a natural starting point as the wave-function overlap between incoming and outgoing modes is expected to be larger for deflections than for forward scattering, due to the network geometry \cite{Qiao2014a,Efimkin2018}. Now, there is only one parameter given by the phase shift $\phi$. For a given orientation, we find that the gapped phases are always AFIs, where the gap is opened via interchannel coupling, without breaking $C_2T$ symmetry. To demonstrate this, we consider the limit $\phi=(n+1/2)\pi$ where the network is localized and the bands are given by degenerate flatbands \cite{Ramires2018}, as illustrated in Fig.\ \ref{fig:2channel}(a). In this case, the network decouples into two versions of the one-channel network with broken $C_2T$ which conserves $C_2T$ on the whole. The decoupled channels are obtained by unitary transformation $\mathcal U = \mathds 1_3 \otimes e^{-i\pi \sigma_y/4} e^{i\phi\sigma_z/2}$, which sends $(a,a') \rightarrow (a_+,a_-)$ with $a_\pm = ( a e^{i \phi / 2} \mp a' e^{-i \phi / 2})/\sqrt{2}$. For the localized network, we then find that $a_+$ ($a_-$) corresponds to the case $P_l=1$ ($P_r=1$) of the one-channel network. From the phase diagram in Fig.\ \ref{fig:phase1}, we see that the $a_+$ modes host a trivial phase, while $a_-$ modes host an AFI phase. For the other valley, the roles of $a_\pm$ are reversed. For general $\phi$, the $a_+$ and $a_-$ modes are coupled. However, as long as the gap is not closed, the AFI phase persists, and we find that this holds for $\pi/6 < \left( \phi \mod \pi \right) < 5\pi/6$.
\begin{figure}
\centering
\includegraphics[width=.9\linewidth]{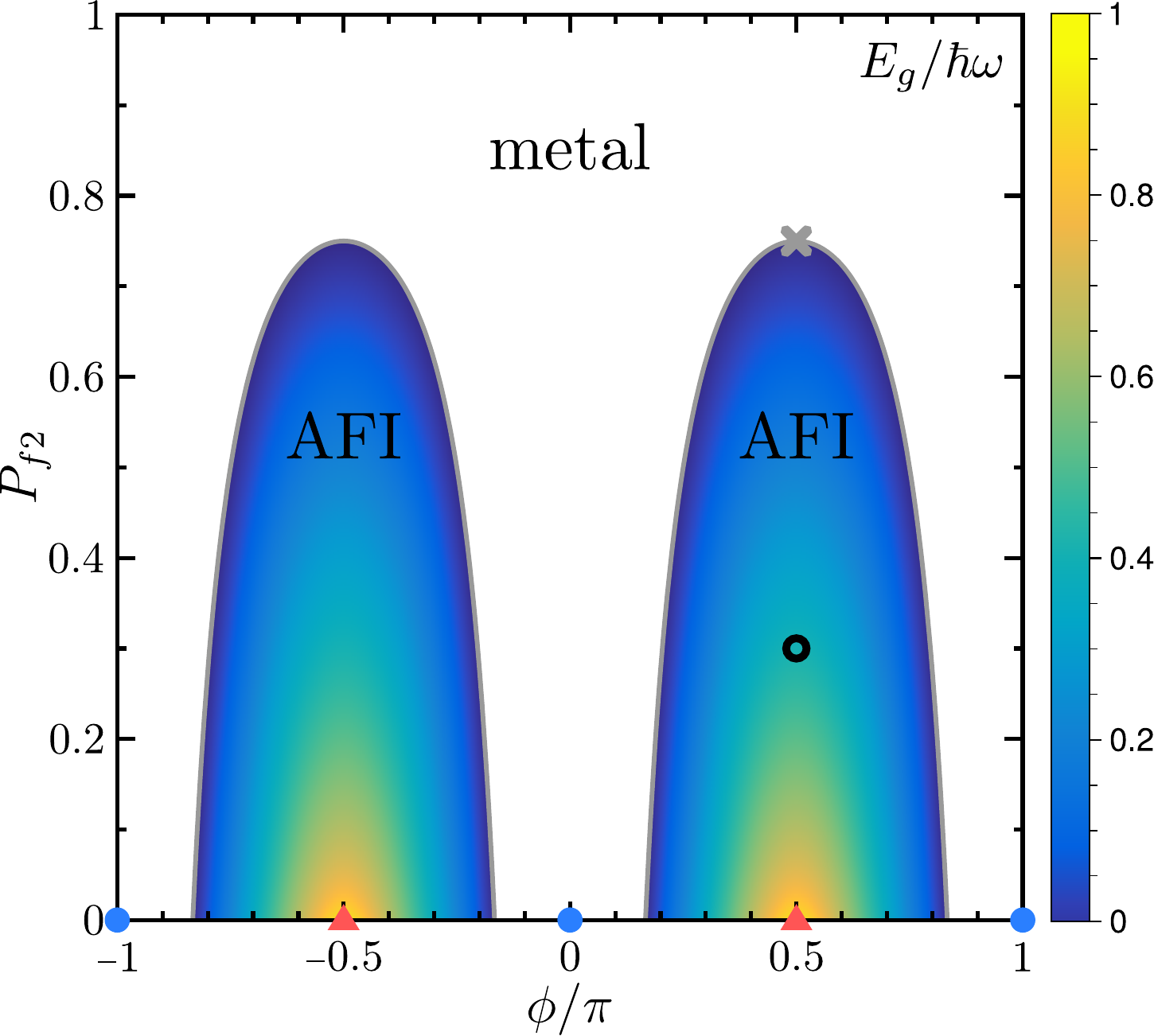}
\caption{Phase diagram for a single valley of the two-channel triangular network with $C_2T$ symmetry with $P_{f1}=0$. In the absence of forward scattering, the AFI phase extends from $\pi/6<|\phi|<5\pi/6$ and the blue dots and red triangles correspond to the chiral zigzag and flatband regime, respectively.}
\label{fig:phase2}
\end{figure}
\begin{figure*}
\centering
\includegraphics[width=\linewidth]{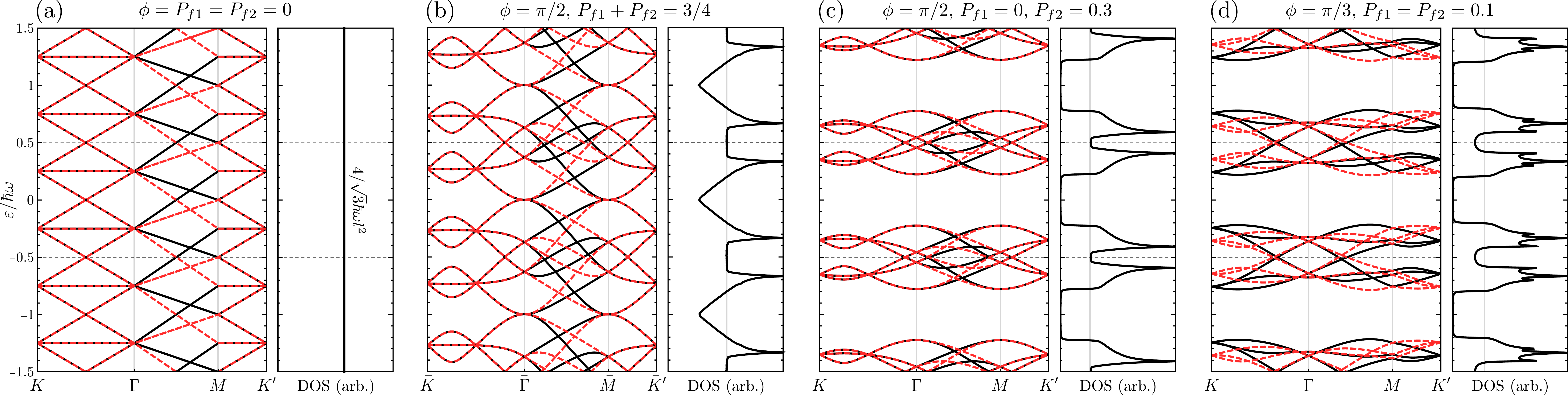}
\caption{Floquet bands for the two-channel case along high-symmetry lines and DOS with $\gamma/\hbar\omega=0.005$. The scattering parameters are shown above the panels and cases (a), (b), and (c) are indicated by the dot, cross, and circle in Fig.\ \ref{fig:phase2}, respectively. Solid (dashed) lines correspond to the (opposite) network orientation [i.e.\ valley index] in Fig.\ \ref{fig:intro}(a).}
\label{fig:bands2}
\end{figure*}

Outside of this range, the network is metallic. Indeed, for $\phi=n\pi$, the network gives rise to three sets of decoupled 1D chiral zigzag (ZZ) modes \cite{Fleischmann2020,Tsim2020,DeBeule2020} that propagate in the $\bm l_j$ ($j=1,2,3$) directions, which is illustrated in Fig.\ \ref{fig:2channel}(b). The ZZ modes have quasienergy bands $\varepsilon_{mj}(\bm k) = \hbar \omega \left( m/2 + 3 \bm k \cdot \bm l_j / 4\pi \right)$, which are shown in Fig.\ \ref{fig:bands2}(a). Moreover, because of their 1D nature and linear dispersion, the DOS of the ZZ modes is constant and equal to $4/\sqrt{3}\hbar \omega l^2$ for a given valley and spin. Note that the velocity of the ZZ modes is half that of the constituent modes, since they traverse twice the direct distance.

\subsection{Effects of forward scattering}

In the limit $P_f = P_{f1}+P_{f2} \rightarrow 1$, it is clear that the network is metallic and therefore forward scattering tends to destroy the AFI phase. The phase boundary where the gap closes can be obtained analytically for $P_{f1}=0$ from the network model \cite{DeBeule2020},
\begin{equation}
\left. P_{f2} \right|_{E_g=0} = 1 - (2\sin \phi)^{-2},
\end{equation}
which is shown in Fig.\ \ref{fig:phase2} together with the gap $E_g$. For $\phi=\pm\pi/2$, the phase boundary always lies at $P_f = 3/4$ as in this case the spectrum depends only on $P_f$. The corresponding Floquet bands and DOS are shown in Fig.\ \ref{fig:bands2}(b). We find that the gap closes both at the $\bar \Gamma$ and $\bar M$ points of the trimer lattice BZ with a quadratic band touching. One might expect that the AFI phase always survives the longest at $\phi = \pm \pi/2$, as in this case the gap attains its maximal value $\hbar \omega$ in the absence of forward scattering. By numerically computing the gap closing points, we find this only holds for $P_{f1}=0$, which is demonstrated in Appendix \ref{app:phase}.

\subsection{Valley anomalous Floquet phase}

We have shown that the AFI phase in the two-channel triangular network with $C_2T$ symmetry can be understood in terms of the one-channel triangular network with broken $C_2T$. However, the main difference lies in the mechanism that opens the gap. While in the one-channel case a gap is only opened when $C_2T$ is broken on the whole, in the two-channel network a gap is opened by interchannel coupling. Secondly, in the two-channel case both valleys simultaneously host anomalous edge states that counterpropagate at a given edge. Hence, with respect to the total winding number it is topologically equivalent to a trivial phase. However, as long as the edge is smooth on the interatomic scale, intervalley scattering is suppressed and the edge hosts a single pair of valley-chiral modes per spin. In the vicinity of the flatband regime, the anomalous edge modes consist mostly of $a_\mp$ modes for valley $K$/$K'$, as these modes host an AFI ($W = \pm1$) phase, while the $a_\pm$ modes host a trivial phase, respectively. It is thus characterized by the valley winding numbers and we refer to it as a valley anomalous Floquet insulator (VAFI) (Table \ref{tab:phases}). In mTBG, however, the type of edge configuration is not immediately obvious and a generic edge will most likely gap out the edge modes of the VAFI.

\begin{figure}
\centering
\includegraphics[width=\linewidth]{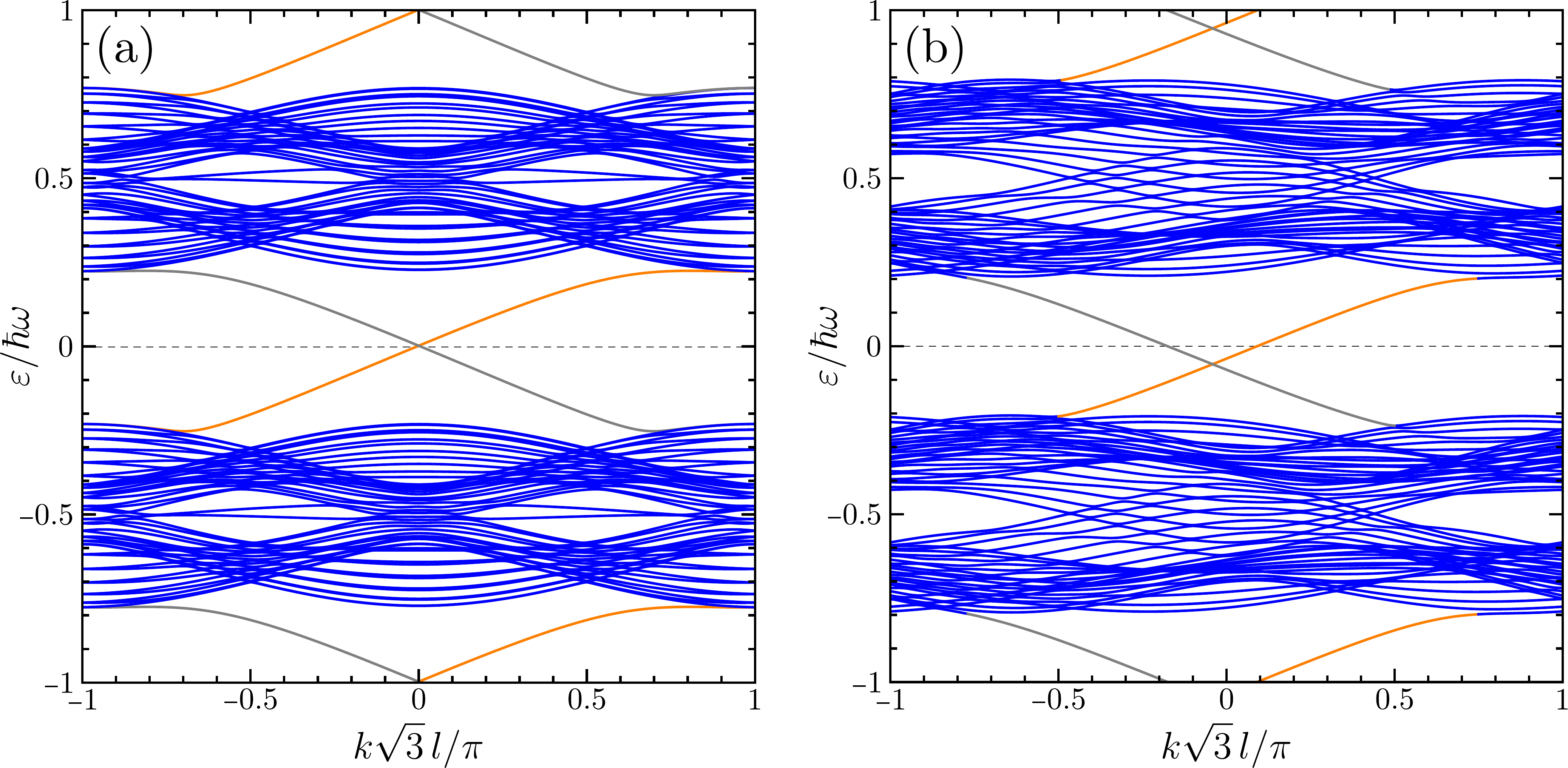} 
\caption{Spectrum of a zigzag ribbon with $N=8$ [Fig.\ \ref{fig:ribbon1}(a)] for the two-channel case with $C_2T$ in the AFI phase for valley $K$ with (a) $\phi=\pi/2$, $P_{f1}=0$, and $P_{f2}=0.3$, and (b) $\phi=\pi/4$ and $P_{f1}=P_{f2}=0.1$. Gray (orange) curves correspond to chiral edge modes localized on the left (right) edge.}
\label{fig:ribbon2}
\end{figure}

Finally, in Fig.\ \ref{fig:ribbon2}, we show the spectrum  in the VAFI phase of a zigzag ribbon of the two-orbital trimer lattice, similar to the system shown in Fig.\ \ref{fig:ribbon1}(a). We note that in Fig.\ \ref{fig:ribbon2}(b) the bulk projection on the zigzag direction does not give a symmetric dispersion with respect to the momentum. However, the symmetry  between opposite momenta along the edge is restored by the other valley so that the total system is time-reversal symmetric. As predicted, we observe chiral edge states at each edge, which mostly consist of $a_-$ modes since these host the VAFI phase in the limits $|\phi| \rightarrow \pi/2$ and $P_f\rightarrow0$. 

\section{Magnetotransport} \label{sec:transport}

\begin{figure}
\centering
\includegraphics[width=\linewidth]{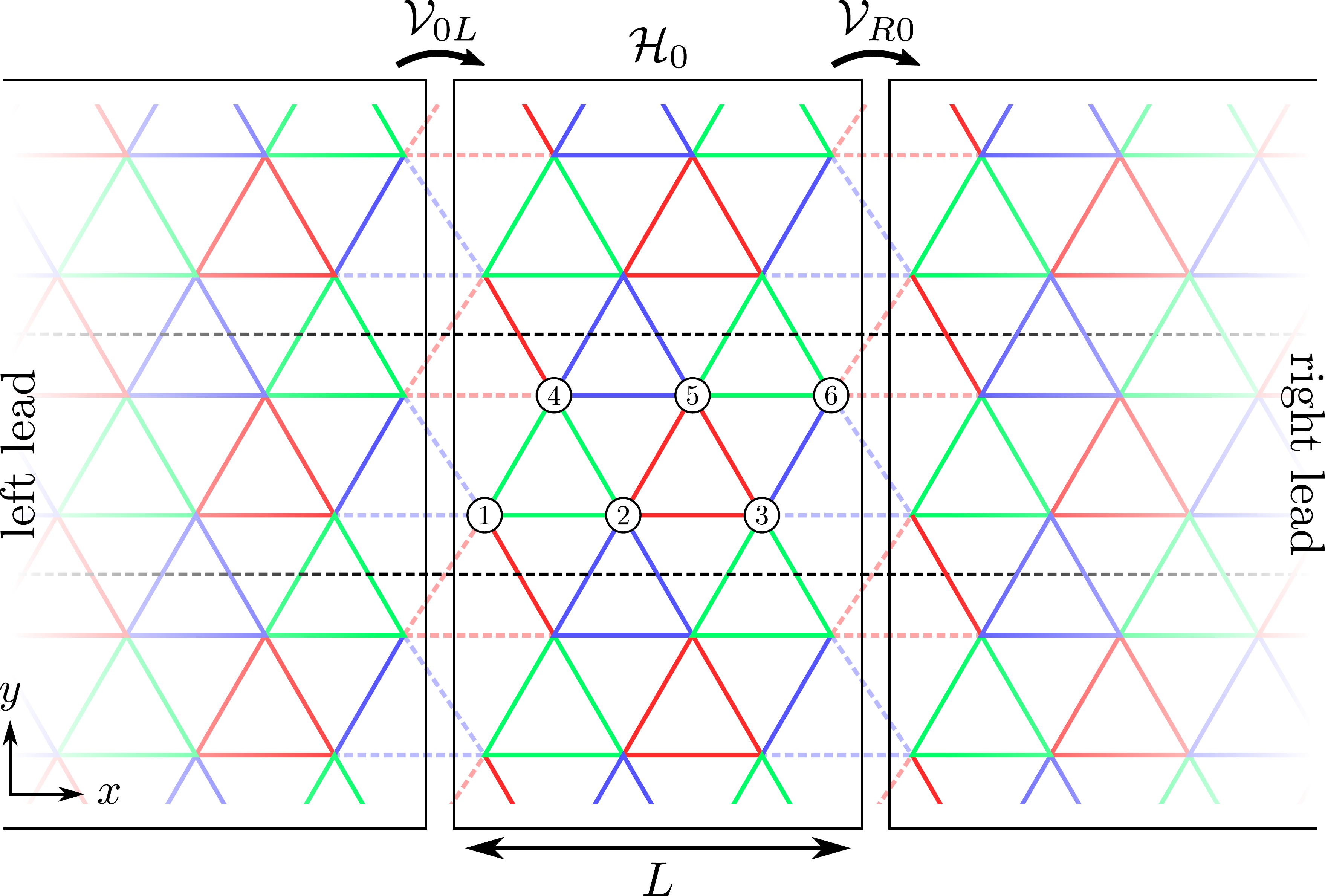}
\caption{Setup for transport calculation where the length of the center region is given by $L=(3N-1/2)l$, shown here for $N=1$. Horizontal dashed lines indicate the unit cell in the transverse direction such that the scattering region with Floquet Hamiltonian $\mathcal H_{0}$ contains a total of $6N$ sites per transverse cell, where the sites are shown as numbered circles.}
\label{fig:transportSetup}
\end{figure}

\begin{figure}
\centering
\includegraphics[width=\linewidth]{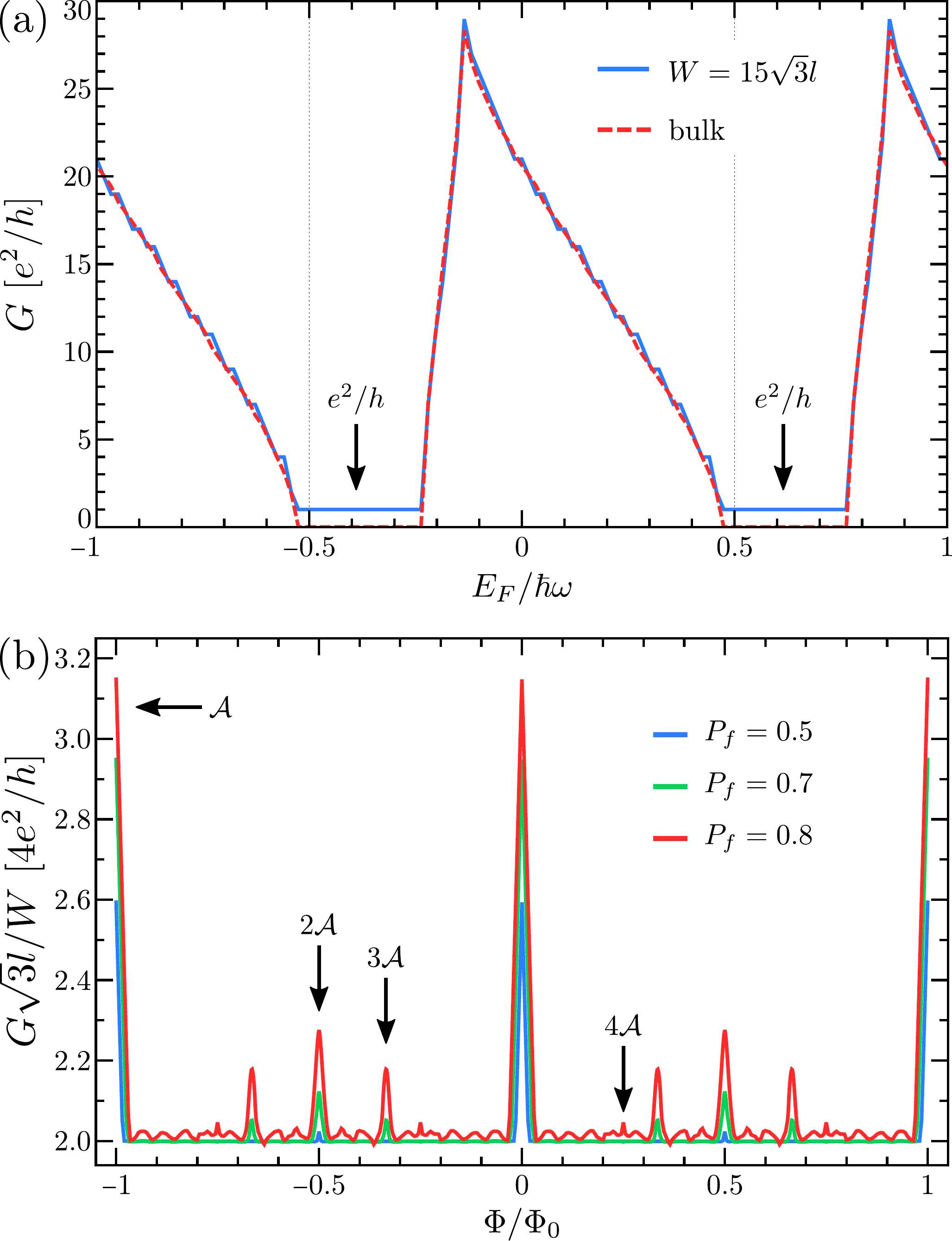}
\label{fig:transportPlot}
\caption{(a) Conductance at zero magnetic field of the one-channel network in the AFI phase with $(P_r,P_l)=(0.6,0.2)$ for an infinitely-long strip of width $W=15\sqrt{3}l$ (solid) and the bulk network (dashed) scaled to match the finite-width result. (b) Magnetoconductance of the two-channel network in mTBG in the ZZ regime ($\phi=0$) for several $P_f$, where $L=8.5l$ and $W\gg L$. Arrows indicate resonances due to paths enclosing a multiple of the moir\'e cell area $\mathcal A = \sqrt{3} \,l^2/2$.}
\end{figure}

To conclude this work, we demonstrate the applicability of the Floquet description by calculating the two-terminal conductance for the setup that is shown in Fig.\ \ref{fig:transportSetup}. The system consists of two semi-infinite leads \cite{LopezSancho1985} and a central scattering region. At zero temperature and in the presence of time-dependent leads \cite{Wu2008}, the time-averaged zero-bias differential conductance reduces to the well-known expression \cite{Moskalets2002, Camalet2003, FoaTorres2014}
\begin{align}
G = \frac{4e^2}{h} \textrm{Tr}_\omega[\Gamma_L G^r_0 \Gamma_R G^a_0],
\end{align}
where $\textrm{Tr}_\omega$ includes a trace over harmonics, which can be thought of as different layers in an effective multilayer system \cite{Wu2008}. Furthermore, $G^a_0 = ( G^r_0 )^\dag = \left( \varepsilon - \mathcal H_{0} - \Sigma_R^a - \Sigma_L^a \right)^{-1}$ is the advanced and retarded Green's function of the system coupled to the right (R) and left (L) leads and $\Gamma_i = 2\textrm{Im} \left[ \Sigma_i^a \right]$ with $i=L,R$ where $\Sigma_i^a = \mathcal V_{0i} g_i^a \mathcal V_{i0}$ is the self energy of the leads. Here, we introduced the Floquet Hamiltonian $\mathcal H_{0}$ of the scattering region, the tunnel couplings $\mathcal V_{0i}$ and $\mathcal V_{i0}$ that couple the scattering region to the leads (Fig.\ \ref{fig:transportSetup}) as well as the surface Green's function $g_i^a$ of the leads.
To illustrate this approach, we calculate the conductance for the one-channel model in the AFI regime. In the presence of a boundary, we find a quantized conductance $e^2/h$ (for a given valley and spin) in the gapped regions, signaling the presence of anomalous edge states [Fig.\ \ref{fig:transportPlot}(a)]. In contrast, the conductance drops to zero for a bulk system with periodic boundary conditions. We want to emphasize that we consider driven leads since they are part of the network. This is in contrast to transport in a driven system with static leads. Therefore there is no Floquet sum rule for edge transport in our setup \cite{Farrell2016}.

We can extend the effective Floquet model to incorporate the effect of a magnetic field $\bm B=B\bm e_z$ on the network. This gives an additional Peierls phase accumulated during propagation between nodes. Here, we assume that the magnetic length is large compared to the scattering region in the network, such that the $S$ matrix of the nodes is not affected by the magnetic field. In this case, the dynamics of the network can be mimicked by introducing new driving steps, namely
\begin{equation}
H(t) = \begin{cases}
H_1, \quad & \quad 0 < t < T/6, \\
H_1', \quad & \quad T/6 < t < T/3, \\
H_2, \quad & \quad T/3 < t < T/2, \\
H_2', \quad & \quad T/2 < t < 2T/3, \\
H_3, \quad & \quad 2T/3 < t < 5T/6, \\
H_3', \quad & \quad 5T/6 < t < T,
\end{cases}
\end{equation}
with $H(t+T)=H(t)$. The additional steps given by $H_1'$, $H_2'$, and $H_3'$ introduce the Peierls phases and are therefore given by on-site terms. For example, in the Landau gauge $\bm A = B(x-l/4)\bm e_y$, the Peierls phase along horizontal links is zero, and given by $\Phi_P(x)$ ($-\Phi_P(x)$) along downward (upward) diagonal links that start at a node with horizontal position $x$, where
\begin{equation}
\Phi_P(x) = \frac{\pi \Phi}{\Phi_0} \frac{x}{l/2},
\end{equation}
with $\Phi = B \sqrt{3}l^2/2$ the flux through a moir\'e cell. In the basis shown in Fig.\ \ref{fig:intro}(d), we thus have
\begin{align}
\frac{H_1'(x)}{\hbar \omega} & = \frac{3\Phi_P(x)}{\pi} \, \textrm{diag} \left( 1 , -1 , 0 \right), \\
\frac{H_2'(x)}{\hbar \omega} & = \frac{3\Phi_P(x)}{\pi} \, \textrm{diag} \left( 0 , 1 , -1 \right), \\
\frac{H_3'(x)}{\hbar \omega} & = \frac{3\Phi_P(x)}{\pi} \, \textrm{diag} \left( -1 , 0 , 1 \right),
\end{align}
where $x$ is the horizontal position of the center of the corresponding trimer. It is straightforward to check that the time-evolution operators indeed give the correct Peierls phases. We have also verified this by explicitly calculating the spectrum of a zigzag ribbon in a magnetic field with both the network and Floquet model, where the $S$ matrices for edge nodes were derived from the Floquet model. It is important to note that this does not correspond to the usual Peierls substitution in the Floquet lattice model. This would introduce an additional flux in the trimers, $\varphi \rightarrow \varphi + \Phi/6$, which corresponds to a different $S$ matrix in the network.
Moreover, the relation between scattering parameters and couplings is modified as each step now lasts for $T/6$ instead of $T/3$. For example, for the one-channel network, we now have
\begin{equation}
\theta_1 = \frac{J}{\hbar \omega} \frac{\pi}{2} \cos \varphi, \qquad \theta_2 = \frac{J}{\hbar \omega} \frac{\pi}{\sqrt{3}} \sin \varphi.
\end{equation}
It is worth mentioning that these extra steps increase the amount of harmonics required to achieve convergence by almost one order of magnitude.

In Fig.\ \ref{fig:transportPlot}(b), we show the conductance as a function of the flux $\Phi/\Phi_0$ for the two-channel network with $\phi=0$ and different $P_f = P_{f1} + P_{f2}$. We find that the magnetoconductance exhibits Aharonov-Bohm resonances whenever an integer amount of flux quanta is threaded through the moir\'e cell. This reproduces previous results obtained with the scattering network approach. We refer to Ref.\ \onlinecite{DeBeule2020} for a detailed discussion on these resonances.

\section{Conclusions} \label{sec:end}

We constructed an effective Floquet lattice model for the oriented triangular scattering network, where the links of the network support either one or two chiral channels. Here, the latter case is realized  in minimally twisted bilayer graphene under interlayer bias. To this end, we first mapped the scattering process at a single node to the dynamics of a three-level system or trimer. The dynamics of the scattering network were then reproduced with a triangular lattice of trimers whose couplings are turned on and off periodically.

We found that the one-channel network hosts a metallic phase and two gapped phases, where the gap is opened by breaking $C_2T$ symmetry. One of the gapped phases is a trivial insulator, while the other is an anomalous Floquet insulator characterized by Floquet winding number $W = \pm 1$ depending on the network orientation. When $C_2$ is conserved but $T$ is broken, the total winding number is finite, while it vanishes when $T$ is conserved. In contrast, in the two-channel network the gap can also be opened by interchannel processes at the nodes without breaking $C_2T$. In this case, each gapped phase corresponds to a valley anomalous Floquet insulator with a pair of counterpropagating anomalous chiral modes at each edge, characterized by a pair of valley winding numbers. This phase has no net winding number and is thus only protected as long as intervalley scattering is absent. The anomalous phase that conserves $C_2T$ is most likely very challenging to realize in minimally twisted bilayer graphene since both experiment and theory indicate that the network remains metallic over a wide range of parameters, e.g., the twist angle or interlayer bias. However, in the presence of external fields that break $C_2T$, one expects an anomalous Floquet phase to arise generically above some critical field strength. The presence of anomalous edge modes could then be probed by non-local transport measurements. Furthermore, the Floquet scheme can in principle also be realized in optical atomic lattices in which case one can tailor the $S$ matrix by controlling the hopping between trimers.

Finally, we performed transport calculations with the effective Floquet model. For the one-channel network, we find a quantized conductance in the AFI phase due to the edge state, while for the two-channel network we reproduced previous results obtained with the network model in the chiral zigzag regime. In particular, we showed that in the presence of forward scattering at the nodes, Aharonov-Bohm oscillations appear in the two-terminal conductance when a magnetic field is applied perpendicularly to the network.

The effective Floquet lattice model that we constructed allows for further research of the topological network in minimally twisted bilayer graphene using standard methods. Moreover, the explicit mapping to the driven system could be interesting for the realization of similar network physics in photonic crystals or optical lattices.

\begin{acknowledgments}
We thank R.\ F.\ Werner for interesting discussions. F.D.\ and P.R.\ gratefully acknowledge funding by the Deutsche Forschungsgemeinschaft (DFG, German Research Foundation) within the framework of Germany's Excellence Strategy -- EXC-2123 QuantumFrontiers -- 390837967.
\end{acknowledgments}

\appendix

\section{Zigzag ribbon} \label{app:ribbon}

The discrete time-dependent Hamiltonian for a zigzag ribbon of the trimer lattice with width $W=(3N-1)l/2$ $(N=1,2,\ldots)$ as illustrated in Fig.\ \ref{fig:ribbon1}(a) for $N=4$, can be written as
\begin{equation} \label{eq:hamZZ}
H(k,t) = \begin{cases}
H_1(k), \quad & \quad 0 < t < T/3, \\ 
H_2(k), \quad & \quad T/3 < t < 2T/3, \\
H_3(k), \quad & \quad 2T/3 < t < T.
\end{cases}
\end{equation}
In the basis given by $\Psi=(\psi_1,\ldots,\psi_{3N})^t$ where the index runs over all trimer sites with labeling defined in Fig.\ \ref{fig:ribbon1}(a), the matrix $H_1$ is block diagonal and the matrices $H_2$ and $H_3$ are block tridiagonal. Explicitly, we have
\begin{equation}
H_j = 
\begin{pmatrix}
H_{0j} & V_j & & \\
V_j^\dag & \ddots & \ddots & \\
& \ddots & \ddots & V_j \\
& & V_j^\dag & H_{0j}
\end{pmatrix},
\end{equation}
for $j=1,2,3$ and where $H_{01} = H_0$ from Eq.\ \eqref{eq:ham0}, $V_1 = 0$, and the remaining matrices are defined below for both the one- and two-channel network. The Floquet Hamiltonian is obtained in the same way as for the bulk.
\begin{figure}
\centering
\includegraphics[width=\linewidth]{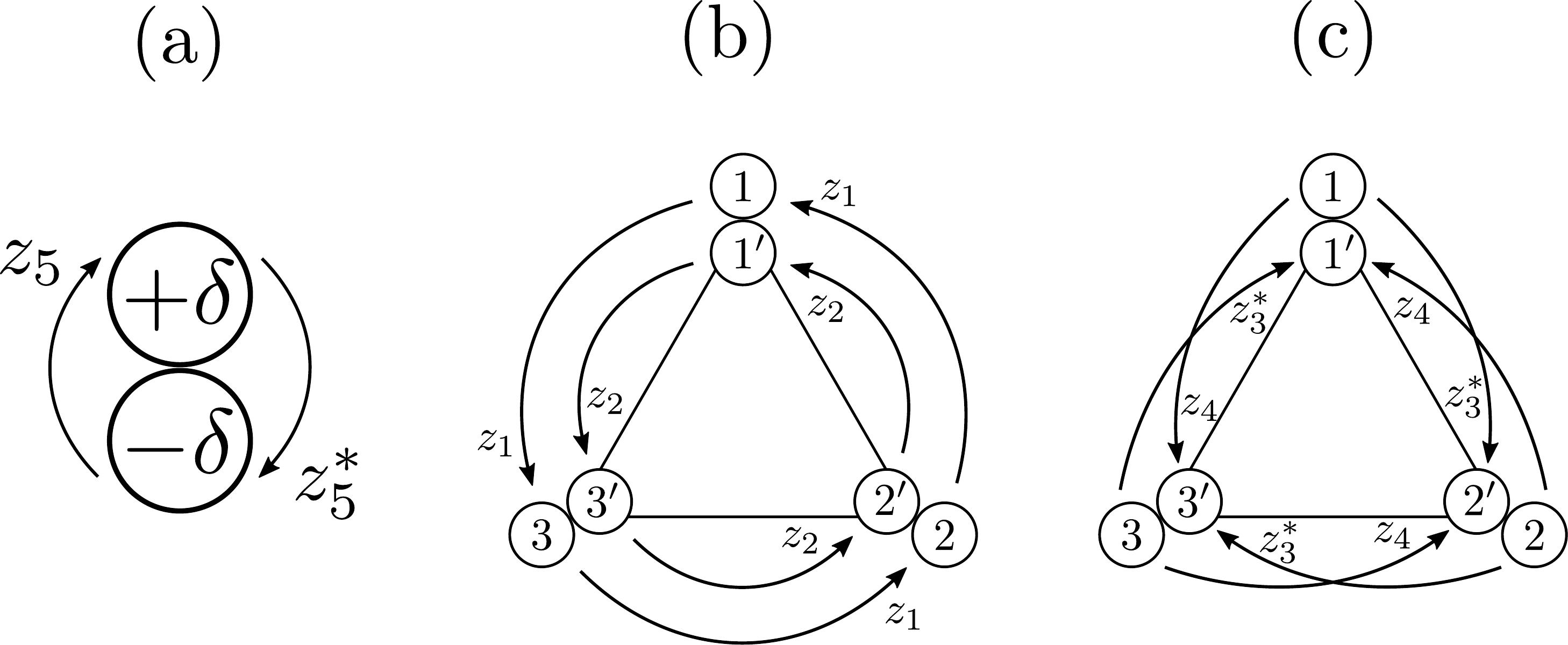}
\caption{Trimer couplings for the two-channel network corresponding to the two-orbital trimer Hamiltonian given in Eq.\ \eqref{eq:ham2} with $z_j=J_je^{i\varphi_j}$. (a) On-site couplings. (b) Intraorbital couplings. (c) Interorbital couplings.}
\label{fig:2coupling}
\end{figure}

\subsection{One-channel network}

In the one-channel case, we have
\begin{equation}
H_{02} =
\begin{pmatrix}
0 & z e^{iq} & 0 \\
z^* e^{-iq} & 0 & 0 \\
0 & 0 & 0
\end{pmatrix}, \,
V_2 =
\begin{pmatrix}
0 & 0 & z^* \\
0 & 0 & z e^{-iq} \\
0 & 0 & 0
\end{pmatrix}, 
\end{equation}
\begin{equation}
H_{03} =
\begin{pmatrix}
0 & 0 & z^* e^{iq} \\
0 & 0 & 0 \\
z e^{-iq} & 0 & 0
\end{pmatrix}, \,
V_3 =
\begin{pmatrix}
0 & 0 & 0 \\
z^* e^{-iq} & 0 & z \\
0 & 0 & 0
\end{pmatrix},
\end{equation}
with $z=Je^{i\varphi}$ and $q = k\sqrt{3}\,l$.

\subsection{Two-channel network}

For the two-channel network, the different couplings are illustrated in Fig.\ \ref{fig:2coupling} and the Hamiltonian matrices for the ribbon become
\begin{equation}
H_{02} =
\begin{pmatrix}
\delta & z_1 e^{iq} & 0 & z_5 & z_3 e^{iq} & 0 \\
z_1^* e^{-iq} & \delta & 0 & z_4^* e^{-iq} & z_5 & 0 \\
0 & 0 & \delta & 0 & 0 & z_5 \\
z_5^* & z_4 e^{iq} & 0 & -\delta & z_2 e^{iq} & 0 \\
z_3^* e^{-iq} & z_5^* & 0 & z_2^* e^{-iq} & -\delta & 0 \\
0 & 0 & z_5^* & 0 & 0 & -\delta
\end{pmatrix},
\end{equation}
\begin{equation}
V_2 =
\begin{pmatrix}
0 & 0 & z_1^* & 0 & 0 & z_4^* \\
0 & 0 & z_1 e^{-iq} & 0 & 0 & z_3 e^{-iq} \\
0 & 0 & 0 & 0 & 0 & 0 \\
0 & 0 & z_3^* & 0 & 0 & z_2^* \\
0 & 0 & z_4 e^{-iq} & 0 & 0 & z_2 e^{-iq} \\
0 & 0 & 0 & 0 & 0 & 0
\end{pmatrix},
\end{equation}
\begin{equation}
H_{03} =
\begin{pmatrix}
\delta & 0 & z_1^* e^{iq} & z_5 & 0 & z_4^* e^{iq} \\
0 & \delta & 0 & 0 & z_5 & 0 \\
z_1 e^{-iq} & 0 & \delta & z_3 e^{-iq} & 0 & z_5 \\
z_5^* & 0 & z_3^* e^{iq} & -\delta & 0 & z_2^* e^{iq}  \\
0 & z_5^* & 0 & 0 & -\delta & 0 \\
z_4 e^{-iq} & 0 & z_5^* & z_2 e^{-iq}  & 0 & -\delta
\end{pmatrix},
\end{equation}
\begin{equation}
V_3 =
\begin{pmatrix}
0 & 0 & 0 & 0 & 0 & 0 \\
z_1^* e^{-iq}  & 0 & z_1 & z_4^* e^{-iq} & 0 & z_3 \\
0 & 0 & 0 & 0 & 0 & 0 \\
0 & 0 & 0 & 0 & 0 & 0 \\
z_3^* e^{-iq} & 0 & z_4 & z_2^* e^{-iq} & 0 & z_2 \\
0 & 0 & 0 & 0 & 0 & 0
\end{pmatrix}.
\end{equation}

\section{Floquet winding number} \label{app:winding}

The bulk topological invariant that characterizes anomalous Floquet insulators is given by the Floquet winding number introduced by Runder \emph{et al.} \cite{Rudner2014}. Here, we calculate the winding number for the one-channel triangular network in the limit where the network is completely localized. To this end, we need the time-evolution operator at all times,
\begin{equation}
\begin{aligned}
& U(\bm k,t) = \\
& \begin{cases}
S(t), \quad & \quad 0 < t < T/3, \\ 
T_1^\dag S \left( t - \tfrac{T}{3} \right) T_1 S_0, \quad & \quad T/3 < t < 2T/3, \\
T_3 S \left( t - \tfrac{2T}{3} \right) T_2 S_0 T_1 S_0, \quad & \quad 2T/3 < t < T,
\end{cases}
\end{aligned}
\end{equation}
with $S(t) = \exp \left( -i H_0 t \right)$, $S_0 = S(T/3)$, and where $H_0$ and $T_i$ ($i=1,2,3$) are defined in Section \ref{sec:1channel}. In case the Floquet operator is trivial, i.e.\ $U(T) = U(0) = 1$, the winding number is defined as
\begin{equation}
\begin{aligned}
W[U] & = \frac{1}{8\pi^2} \int_0^T dt \int_\textrm{BZ} d^2 \bm k \\
& \textrm{Tr} \left( U^{-1} \left( \partial_t U \right) \left[ U^{-1} \left( \partial_{k_x} U \right), U^{-1} \left( \partial_{k_y} U \right) \right] \right),
\end{aligned}
\end{equation}
where the momentum integral runs over the first Brillouin zone. For the gapped one-channel network, the Floquet operator is periodic for $P_l = 1$ or $P_r = 1$. In the former case, for the orientation shown in Fig.\ \ref{fig:intro}, $S(t) = 1$, so that $U (t)= 1$ is trivial at all times and $W=0$. In the latter case, we have $J/\hbar \omega=1/\sqrt{3}$ and $\varphi=\pi/2$. We find that the trace is independent of $\bm k$ and
\begin{equation}
\begin{aligned}
& W[P_r=1] = \frac{l^2 \Omega_\textrm{BZ}}{8\pi^2} \\
& \quad \times 8 \pi \int_0^{1/3} ds \sin \left( \pi s + \tfrac{\pi}{3} \right) \sin \left( 3\pi s \right),
\end{aligned}
\end{equation}
with $s=t/T$, $l^2 \Omega_\textrm{BZ} = 8\pi^2/3\sqrt{3}$, and where the last factor evaluates to $3\sqrt{3}$. Hence, we obtain $W[P_l=1]=0$ and $W[P_r=1]=1$. If the orientation (valley) is reversed, $W[P_l'=1]=-1$ and $W[P_r'=1]=0$, where the prime indicates different orientation. Moreover, when $C_2$ is conserved but $T$ broken, $P_{r,l}' = P_{r,l}$ and $W = \sgn \left( P_r - P_l \right)$ where we used the fact that the winding number does not change as long as the energy gap does not close. On the other hand, when $T$ is conserved but $C_2$ is broken, we have $P_{r,l}' = P_{l,r}$ and the total winding number vanishes. However, in the absence of intervalley processes, one can still define two valley winding numbers that can be nonzero (Table \ref{tab:phases}). When both of these symmetries are absent, there is no relation between scattering parameters of opposite valleys and in general, we have
\begin{equation}
W = \theta \left( P_r - P_l \right) - \theta \left( P_l' - P_r' \right).
\end{equation}

\section{Phase diagram} \label{app:phase}

\begin{figure}
\centering
\includegraphics[width=.85\linewidth]{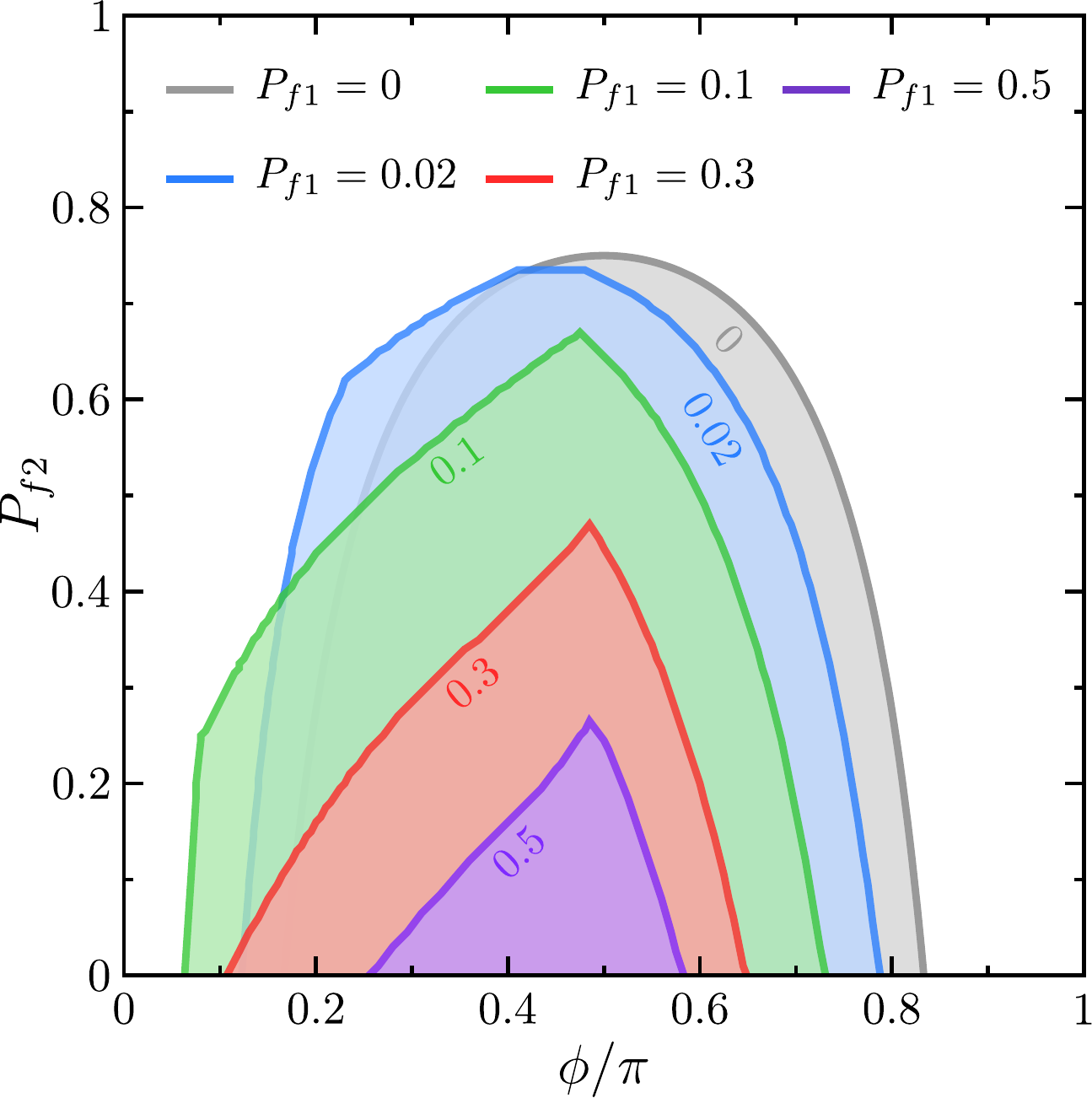}
\caption{Phase diagram of the two-channel triangular network with $C_2T$ symmetry. Colored regions correspond to the AFI phase, while white regions are metallic.}
\label{fig:phase2_Pf1}
\end{figure}

The phase diagram of the two-channel network with $C_2T$ symmetry and $P_{d1} = P_{d2}$ in the $(\phi,P_{f2})$ plane was obtained by numerically computing the gap-closing points. This is shown for several values of $P_{f1}$ in Fig.\ \ref{fig:phase2_Pf1}.

\section{Transport} \label{app:transport}

We calculated transport in the $\bm l_3$ direction of the trimer lattice. To this end, we use the setup shown in Fig.\ \ref{fig:transportSetup} where the transport direction is given by the $x$ axis. Here, we take a unit cell made up of two rows of sites, which can be subdivided into sets of six sites as shown in the figure. The Hamiltonian for each step then has the same structure as Eq.\ \eqref{eq:hamZZ}, where each block is now given by a $6\times6$ ($12\times12$) matrix for the one(two)-channel case. For the one-channel network, we obtain
\begin{equation}
H_{01} =
\begin{pmatrix}
0 & z^* & 0 & z & 0 & 0 \\
z & 0 & 0 & z^* & 0 & 0 \\
0 & 0 & 0 & 0 & z^* e^{-iq} & z e^{-iq} \\
z^* & z & 0 & 0 & 0 & 0 \\
0 & 0 & z e^{iq} & 0 & 0 & z^* \\
0 & 0 & z^* e^{iq} & 0 & z & 0 
\end{pmatrix},
\end{equation}
and $V_1=0$, while
\begin{equation}
H_{02} =
\begin{pmatrix}
0 & 0 & 0 & 0 & 0 & 0 \\
0 & 0 & 0 & z^* e^{-iq} & z e^{-iq} & 0 \\
0 & 0 & 0 & 0 & 0 & z \\
0 & z e^{iq} & 0 & 0 & z^* & 0 \\
0 & z^* e^{iq} & 0 & z & 0 & 0 \\
0 & 0 & z^* & 0 & 0 & 0 
\end{pmatrix},
\end{equation}
and $[V_2]_{mn} = z \delta_{m3} \delta_{1n} + z^* \delta_{m6} \delta_{1n}$. Finally, we have
\begin{equation}
H_{03} =
\begin{pmatrix}
0 & 0 & 0 & z e^{-iq} & 0 & 0 \\
0 & 0 & z^* & 0 & z & 0 \\
0 & z & 0 & 0 & z^* & 0 \\
z^* e^{iq} & 0 & 0 & 0 & 0 & 0 \\
0 & z^* & z & 0 & 0 & 0 \\
0 & 0 & 0 & 0 & 0 & 0 
\end{pmatrix},
\end{equation}
and $[V_3]_{mn} = z^* \delta_{m6} \delta_{4n} + z e^{iq} \delta_{m6} \delta_{1n}$. The matrices for the two-channel case can be obtained similarly.

\clearpage
\bibliography{references}

\begin{thebibliography}{60}%
\makeatletter
\providecommand \@ifxundefined [1]{%
 \@ifx{#1\undefined}
}%
\providecommand \@ifnum [1]{%
 \ifnum #1\expandafter \@firstoftwo
 \else \expandafter \@secondoftwo
 \fi
}%
\providecommand \@ifx [1]{%
 \ifx #1\expandafter \@firstoftwo
 \else \expandafter \@secondoftwo
 \fi
}%
\providecommand \natexlab [1]{#1}%
\providecommand \enquote  [1]{``#1''}%
\providecommand \bibnamefont  [1]{#1}%
\providecommand \bibfnamefont [1]{#1}%
\providecommand \citenamefont [1]{#1}%
\providecommand \href@noop [0]{\@secondoftwo}%
\providecommand \href [0]{\begingroup \@sanitize@url \@href}%
\providecommand \@href[1]{\@@startlink{#1}\@@href}%
\providecommand \@@href[1]{\endgroup#1\@@endlink}%
\providecommand \@sanitize@url [0]{\catcode `\\12\catcode `\$12\catcode
  `\&12\catcode `\#12\catcode `\^12\catcode `\_12\catcode `\%12\relax}%
\providecommand \@@startlink[1]{}%
\providecommand \@@endlink[0]{}%
\providecommand \url  [0]{\begingroup\@sanitize@url \@url }%
\providecommand \@url [1]{\endgroup\@href {#1}{\urlprefix }}%
\providecommand \urlprefix  [0]{URL }%
\providecommand \Eprint [0]{\href }%
\providecommand \doibase [0]{http://dx.doi.org/}%
\providecommand \selectlanguage [0]{\@gobble}%
\providecommand \bibinfo  [0]{\@secondoftwo}%
\providecommand \bibfield  [0]{\@secondoftwo}%
\providecommand \translation [1]{[#1]}%
\providecommand \BibitemOpen [0]{}%
\providecommand \bibitemStop [0]{}%
\providecommand \bibitemNoStop [0]{.\EOS\space}%
\providecommand \EOS [0]{\spacefactor3000\relax}%
\providecommand \BibitemShut  [1]{\csname bibitem#1\endcsname}%
\let\auto@bib@innerbib\@empty
\bibitem [{\citenamefont {{Lopes dos Santos}}\ \emph
  {et~al.}(2007)\citenamefont {{Lopes dos Santos}}, \citenamefont {Peres},\
  and\ \citenamefont {{Castro Neto}}}]{LopesDosSantos2007}%
  \BibitemOpen
  \bibfield  {author} {\bibinfo {author} {\bibfnamefont {J.~M.~B.}\
  \bibnamefont {{Lopes dos Santos}}}, \bibinfo {author} {\bibfnamefont
  {N.~M.~R.}\ \bibnamefont {Peres}}, \ and\ \bibinfo {author} {\bibfnamefont
  {A.~H.}\ \bibnamefont {{Castro Neto}}},\ }\href {\doibase
  10.1103/PhysRevLett.99.256802} {\bibfield  {journal} {\bibinfo  {journal}
  {Phys. Rev. Lett.}\ }\textbf {\bibinfo {volume} {99}},\ \bibinfo {pages}
  {256802} (\bibinfo {year} {2007})}\BibitemShut {NoStop}%
\bibitem [{\citenamefont {{Su{\'{a}}rez Morell}}\ \emph
  {et~al.}(2010)\citenamefont {{Su{\'{a}}rez Morell}}, \citenamefont {Correa},
  \citenamefont {Vargas}, \citenamefont {Pacheco},\ and\ \citenamefont
  {Barticevic}}]{SuarezMorell2010}%
  \BibitemOpen
  \bibfield  {author} {\bibinfo {author} {\bibfnamefont {E.}~\bibnamefont
  {{Su{\'{a}}rez Morell}}}, \bibinfo {author} {\bibfnamefont {J.~D.}\
  \bibnamefont {Correa}}, \bibinfo {author} {\bibfnamefont {P.}~\bibnamefont
  {Vargas}}, \bibinfo {author} {\bibfnamefont {M.}~\bibnamefont {Pacheco}}, \
  and\ \bibinfo {author} {\bibfnamefont {Z.}~\bibnamefont {Barticevic}},\
  }\href {\doibase 10.1103/PhysRevB.82.121407} {\bibfield  {journal} {\bibinfo
  {journal} {Phys. Rev. B}\ }\textbf {\bibinfo {volume} {82}},\ \bibinfo
  {pages} {121407(R)} (\bibinfo {year} {2010})}\BibitemShut {NoStop}%
\bibitem [{\citenamefont {Bistritzer}\ and\ \citenamefont
  {MacDonald}(2011)}]{Bistritzer2010}%
  \BibitemOpen
  \bibfield  {author} {\bibinfo {author} {\bibfnamefont {R.}~\bibnamefont
  {Bistritzer}}\ and\ \bibinfo {author} {\bibfnamefont {A.~H.}\ \bibnamefont
  {MacDonald}},\ }\href {\doibase 10.1073/pnas.1108174108} {\bibfield
  {journal} {\bibinfo  {journal} {Proc. Natl. Acad. Sci.}\ }\textbf {\bibinfo
  {volume} {108}},\ \bibinfo {pages} {12233} (\bibinfo {year}
  {2011})}\BibitemShut {NoStop}%
\bibitem [{\citenamefont {Li}\ \emph {et~al.}(2010)\citenamefont {Li},
  \citenamefont {Luican}, \citenamefont {{Lopes dos Santos}}, \citenamefont
  {{Castro Neto}}, \citenamefont {Reina}, \citenamefont {Kong},\ and\
  \citenamefont {Andrei}}]{Li2010}%
  \BibitemOpen
  \bibfield  {author} {\bibinfo {author} {\bibfnamefont {G.}~\bibnamefont
  {Li}}, \bibinfo {author} {\bibfnamefont {A.}~\bibnamefont {Luican}}, \bibinfo
  {author} {\bibfnamefont {J.~M.~B.}\ \bibnamefont {{Lopes dos Santos}}},
  \bibinfo {author} {\bibfnamefont {A.~H.}\ \bibnamefont {{Castro Neto}}},
  \bibinfo {author} {\bibfnamefont {A.}~\bibnamefont {Reina}}, \bibinfo
  {author} {\bibfnamefont {J.}~\bibnamefont {Kong}}, \ and\ \bibinfo {author}
  {\bibfnamefont {E.~Y.}\ \bibnamefont {Andrei}},\ }\href {\doibase
  10.1038/nphys1463} {\bibfield  {journal} {\bibinfo  {journal} {Nat. Phys.}\
  }\textbf {\bibinfo {volume} {6}},\ \bibinfo {pages} {109} (\bibinfo {year}
  {2010})}\BibitemShut {NoStop}%
\bibitem [{\citenamefont {Kim}\ \emph {et~al.}(2017)\citenamefont {Kim},
  \citenamefont {DaSilva}, \citenamefont {Huang}, \citenamefont {Fallahazad},
  \citenamefont {Larentis}, \citenamefont {Taniguchi}, \citenamefont
  {Watanabe}, \citenamefont {LeRoy}, \citenamefont {MacDonald},\ and\
  \citenamefont {Tutuc}}]{Kim2017a}%
  \BibitemOpen
  \bibfield  {author} {\bibinfo {author} {\bibfnamefont {K.}~\bibnamefont
  {Kim}}, \bibinfo {author} {\bibfnamefont {A.}~\bibnamefont {DaSilva}},
  \bibinfo {author} {\bibfnamefont {S.}~\bibnamefont {Huang}}, \bibinfo
  {author} {\bibfnamefont {B.}~\bibnamefont {Fallahazad}}, \bibinfo {author}
  {\bibfnamefont {S.}~\bibnamefont {Larentis}}, \bibinfo {author}
  {\bibfnamefont {T.}~\bibnamefont {Taniguchi}}, \bibinfo {author}
  {\bibfnamefont {K.}~\bibnamefont {Watanabe}}, \bibinfo {author}
  {\bibfnamefont {B.~J.}\ \bibnamefont {LeRoy}}, \bibinfo {author}
  {\bibfnamefont {A.~H.}\ \bibnamefont {MacDonald}}, \ and\ \bibinfo {author}
  {\bibfnamefont {E.}~\bibnamefont {Tutuc}},\ }\href {\doibase
  10.1073/pnas.1620140114} {\bibfield  {journal} {\bibinfo  {journal} {Proc.
  Natl. Acad. Sci.}\ }\textbf {\bibinfo {volume} {114}},\ \bibinfo {pages}
  {3364} (\bibinfo {year} {2017})}\BibitemShut {NoStop}%
\bibitem [{\citenamefont {Cao}\ \emph {et~al.}(2018{\natexlab{a}})\citenamefont
  {Cao}, \citenamefont {Fatemi}, \citenamefont {Demir}, \citenamefont {Fang},
  \citenamefont {Tomarken}, \citenamefont {Luo}, \citenamefont
  {Sanchez-Yamagishi}, \citenamefont {Watanabe}, \citenamefont {Taniguchi},
  \citenamefont {Kaxiras}, \citenamefont {Ashoori},\ and\ \citenamefont
  {Jarillo-Herrero}}]{Cao2018a}%
  \BibitemOpen
  \bibfield  {author} {\bibinfo {author} {\bibfnamefont {Y.}~\bibnamefont
  {Cao}}, \bibinfo {author} {\bibfnamefont {V.}~\bibnamefont {Fatemi}},
  \bibinfo {author} {\bibfnamefont {A.}~\bibnamefont {Demir}}, \bibinfo
  {author} {\bibfnamefont {S.}~\bibnamefont {Fang}}, \bibinfo {author}
  {\bibfnamefont {S.~L.}\ \bibnamefont {Tomarken}}, \bibinfo {author}
  {\bibfnamefont {J.~Y.}\ \bibnamefont {Luo}}, \bibinfo {author} {\bibfnamefont
  {J.~D.}\ \bibnamefont {Sanchez-Yamagishi}}, \bibinfo {author} {\bibfnamefont
  {K.}~\bibnamefont {Watanabe}}, \bibinfo {author} {\bibfnamefont
  {T.}~\bibnamefont {Taniguchi}}, \bibinfo {author} {\bibfnamefont
  {E.}~\bibnamefont {Kaxiras}}, \bibinfo {author} {\bibfnamefont {R.~C.}\
  \bibnamefont {Ashoori}}, \ and\ \bibinfo {author} {\bibfnamefont
  {P.}~\bibnamefont {Jarillo-Herrero}},\ }\href {\doibase 10.1038/nature26154}
  {\bibfield  {journal} {\bibinfo  {journal} {Nature}\ }\textbf {\bibinfo
  {volume} {556}},\ \bibinfo {pages} {80} (\bibinfo {year}
  {2018}{\natexlab{a}})}\BibitemShut {NoStop}%
\bibitem [{\citenamefont {Cao}\ \emph {et~al.}(2018{\natexlab{b}})\citenamefont
  {Cao}, \citenamefont {Fatemi}, \citenamefont {Fang}, \citenamefont
  {Watanabe}, \citenamefont {Taniguchi}, \citenamefont {Kaxiras},\ and\
  \citenamefont {Jarillo-Herrero}}]{Cao2018}%
  \BibitemOpen
  \bibfield  {author} {\bibinfo {author} {\bibfnamefont {Y.}~\bibnamefont
  {Cao}}, \bibinfo {author} {\bibfnamefont {V.}~\bibnamefont {Fatemi}},
  \bibinfo {author} {\bibfnamefont {S.}~\bibnamefont {Fang}}, \bibinfo {author}
  {\bibfnamefont {K.}~\bibnamefont {Watanabe}}, \bibinfo {author}
  {\bibfnamefont {T.}~\bibnamefont {Taniguchi}}, \bibinfo {author}
  {\bibfnamefont {E.}~\bibnamefont {Kaxiras}}, \ and\ \bibinfo {author}
  {\bibfnamefont {P.}~\bibnamefont {Jarillo-Herrero}},\ }\href {\doibase
  10.1038/nature26160} {\bibfield  {journal} {\bibinfo  {journal} {Nature}\
  }\textbf {\bibinfo {volume} {556}},\ \bibinfo {pages} {43} (\bibinfo {year}
  {2018}{\natexlab{b}})}\BibitemShut {NoStop}%
\bibitem [{\citenamefont {Yankowitz}\ \emph {et~al.}(2019)\citenamefont
  {Yankowitz}, \citenamefont {Chen}, \citenamefont {Polshyn}, \citenamefont
  {Zhang}, \citenamefont {Watanabe}, \citenamefont {Taniguchi}, \citenamefont
  {Graf}, \citenamefont {Young},\ and\ \citenamefont {Dean}}]{Yankowitz2019}%
  \BibitemOpen
  \bibfield  {author} {\bibinfo {author} {\bibfnamefont {M.}~\bibnamefont
  {Yankowitz}}, \bibinfo {author} {\bibfnamefont {S.}~\bibnamefont {Chen}},
  \bibinfo {author} {\bibfnamefont {H.}~\bibnamefont {Polshyn}}, \bibinfo
  {author} {\bibfnamefont {Y.}~\bibnamefont {Zhang}}, \bibinfo {author}
  {\bibfnamefont {K.}~\bibnamefont {Watanabe}}, \bibinfo {author}
  {\bibfnamefont {T.}~\bibnamefont {Taniguchi}}, \bibinfo {author}
  {\bibfnamefont {D.}~\bibnamefont {Graf}}, \bibinfo {author} {\bibfnamefont
  {A.~F.}\ \bibnamefont {Young}}, \ and\ \bibinfo {author} {\bibfnamefont
  {C.~R.}\ \bibnamefont {Dean}},\ }\href {\doibase 10.1126/science.aav1910}
  {\bibfield  {journal} {\bibinfo  {journal} {Science}\ }\textbf {\bibinfo
  {volume} {363}},\ \bibinfo {pages} {1059} (\bibinfo {year}
  {2019})}\BibitemShut {NoStop}%
\bibitem [{\citenamefont {Sharpe}\ \emph {et~al.}(2019)\citenamefont {Sharpe},
  \citenamefont {Fox}, \citenamefont {Barnard}, \citenamefont {Finney},
  \citenamefont {Watanabe}, \citenamefont {Taniguchi}, \citenamefont
  {Kastner},\ and\ \citenamefont {Goldhaber-Gordon}}]{Sharpe2019}%
  \BibitemOpen
  \bibfield  {author} {\bibinfo {author} {\bibfnamefont {A.~L.}\ \bibnamefont
  {Sharpe}}, \bibinfo {author} {\bibfnamefont {E.~J.}\ \bibnamefont {Fox}},
  \bibinfo {author} {\bibfnamefont {A.~W.}\ \bibnamefont {Barnard}}, \bibinfo
  {author} {\bibfnamefont {J.}~\bibnamefont {Finney}}, \bibinfo {author}
  {\bibfnamefont {K.}~\bibnamefont {Watanabe}}, \bibinfo {author}
  {\bibfnamefont {T.}~\bibnamefont {Taniguchi}}, \bibinfo {author}
  {\bibfnamefont {M.~A.}\ \bibnamefont {Kastner}}, \ and\ \bibinfo {author}
  {\bibfnamefont {D.}~\bibnamefont {Goldhaber-Gordon}},\ }\href {\doibase
  10.1126/science.aaw3780} {\bibfield  {journal} {\bibinfo  {journal}
  {Science}\ }\textbf {\bibinfo {volume} {365}},\ \bibinfo {pages} {605}
  (\bibinfo {year} {2019})}\BibitemShut {NoStop}%
\bibitem [{\citenamefont {Kerelsky}\ \emph {et~al.}(2019)\citenamefont
  {Kerelsky}, \citenamefont {McGilly}, \citenamefont {Kennes}, \citenamefont
  {Xian}, \citenamefont {Yankowitz}, \citenamefont {Chen}, \citenamefont
  {Watanabe}, \citenamefont {Taniguchi}, \citenamefont {Hone}, \citenamefont
  {Dean}, \citenamefont {Rubio},\ and\ \citenamefont
  {Pasupathy}}]{Kerelsky2019}%
  \BibitemOpen
  \bibfield  {author} {\bibinfo {author} {\bibfnamefont {A.}~\bibnamefont
  {Kerelsky}}, \bibinfo {author} {\bibfnamefont {L.~J.}\ \bibnamefont
  {McGilly}}, \bibinfo {author} {\bibfnamefont {D.~M.}\ \bibnamefont {Kennes}},
  \bibinfo {author} {\bibfnamefont {L.}~\bibnamefont {Xian}}, \bibinfo {author}
  {\bibfnamefont {M.}~\bibnamefont {Yankowitz}}, \bibinfo {author}
  {\bibfnamefont {S.}~\bibnamefont {Chen}}, \bibinfo {author} {\bibfnamefont
  {K.}~\bibnamefont {Watanabe}}, \bibinfo {author} {\bibfnamefont
  {T.}~\bibnamefont {Taniguchi}}, \bibinfo {author} {\bibfnamefont
  {J.}~\bibnamefont {Hone}}, \bibinfo {author} {\bibfnamefont {C.}~\bibnamefont
  {Dean}}, \bibinfo {author} {\bibfnamefont {A.}~\bibnamefont {Rubio}}, \ and\
  \bibinfo {author} {\bibfnamefont {A.~N.}\ \bibnamefont {Pasupathy}},\ }\href
  {\doibase 10.1038/s41586-019-1431-9} {\bibfield  {journal} {\bibinfo
  {journal} {Nature}\ }\textbf {\bibinfo {volume} {572}},\ \bibinfo {pages}
  {95} (\bibinfo {year} {2019})}\BibitemShut {NoStop}%
\bibitem [{\citenamefont {Choi}\ \emph {et~al.}(2019)\citenamefont {Choi},
  \citenamefont {Kemmer}, \citenamefont {Peng}, \citenamefont {Thomson},
  \citenamefont {Arora}, \citenamefont {Polski}, \citenamefont {Zhang},
  \citenamefont {Ren}, \citenamefont {Alicea}, \citenamefont {Refael},
  \citenamefont {von Oppen}, \citenamefont {Watanabe}, \citenamefont
  {Taniguchi},\ and\ \citenamefont {Nadj-Perge}}]{Choi2019}%
  \BibitemOpen
  \bibfield  {author} {\bibinfo {author} {\bibfnamefont {Y.}~\bibnamefont
  {Choi}}, \bibinfo {author} {\bibfnamefont {J.}~\bibnamefont {Kemmer}},
  \bibinfo {author} {\bibfnamefont {Y.}~\bibnamefont {Peng}}, \bibinfo {author}
  {\bibfnamefont {A.}~\bibnamefont {Thomson}}, \bibinfo {author} {\bibfnamefont
  {H.}~\bibnamefont {Arora}}, \bibinfo {author} {\bibfnamefont
  {R.}~\bibnamefont {Polski}}, \bibinfo {author} {\bibfnamefont
  {Y.}~\bibnamefont {Zhang}}, \bibinfo {author} {\bibfnamefont
  {H.}~\bibnamefont {Ren}}, \bibinfo {author} {\bibfnamefont {J.}~\bibnamefont
  {Alicea}}, \bibinfo {author} {\bibfnamefont {G.}~\bibnamefont {Refael}},
  \bibinfo {author} {\bibfnamefont {F.}~\bibnamefont {von Oppen}}, \bibinfo
  {author} {\bibfnamefont {K.}~\bibnamefont {Watanabe}}, \bibinfo {author}
  {\bibfnamefont {T.}~\bibnamefont {Taniguchi}}, \ and\ \bibinfo {author}
  {\bibfnamefont {S.}~\bibnamefont {Nadj-Perge}},\ }\href {\doibase
  10.1038/s41567-019-0606-5} {\bibfield  {journal} {\bibinfo  {journal} {Nat.
  Phys.}\ }\textbf {\bibinfo {volume} {15}},\ \bibinfo {pages} {1174} (\bibinfo
  {year} {2019})}\BibitemShut {NoStop}%
\bibitem [{\citenamefont {Cao}\ \emph {et~al.}(2020)\citenamefont {Cao},
  \citenamefont {Chowdhury}, \citenamefont {Rodan-Legrain}, \citenamefont
  {Rubies-Bigord{\`{a}}}, \citenamefont {Watanabe}, \citenamefont {Taniguchi},
  \citenamefont {Senthil},\ and\ \citenamefont {Jarillo-Herrero}}]{Cao2019}%
  \BibitemOpen
  \bibfield  {author} {\bibinfo {author} {\bibfnamefont {Y.}~\bibnamefont
  {Cao}}, \bibinfo {author} {\bibfnamefont {D.}~\bibnamefont {Chowdhury}},
  \bibinfo {author} {\bibfnamefont {D.}~\bibnamefont {Rodan-Legrain}}, \bibinfo
  {author} {\bibfnamefont {O.}~\bibnamefont {Rubies-Bigord{\`{a}}}}, \bibinfo
  {author} {\bibfnamefont {K.}~\bibnamefont {Watanabe}}, \bibinfo {author}
  {\bibfnamefont {T.}~\bibnamefont {Taniguchi}}, \bibinfo {author}
  {\bibfnamefont {T.}~\bibnamefont {Senthil}}, \ and\ \bibinfo {author}
  {\bibfnamefont {P.}~\bibnamefont {Jarillo-Herrero}},\ }\href {\doibase
  10.1103/PhysRevLett.124.076801} {\bibfield  {journal} {\bibinfo  {journal}
  {Phys. Rev. Lett.}\ }\textbf {\bibinfo {volume} {124}},\ \bibinfo {pages}
  {076801} (\bibinfo {year} {2020})}\BibitemShut {NoStop}%
\bibitem [{\citenamefont {Nam}\ and\ \citenamefont {Koshino}(2017)}]{Nam2017}%
  \BibitemOpen
  \bibfield  {author} {\bibinfo {author} {\bibfnamefont {N.~N.~T.}\
  \bibnamefont {Nam}}\ and\ \bibinfo {author} {\bibfnamefont {M.}~\bibnamefont
  {Koshino}},\ }\href {\doibase 10.1103/PhysRevB.96.075311} {\bibfield
  {journal} {\bibinfo  {journal} {Phys. Rev. B}\ }\textbf {\bibinfo {volume}
  {96}},\ \bibinfo {pages} {075311} (\bibinfo {year} {2017})}\BibitemShut
  {NoStop}%
\bibitem [{\citenamefont {Yoo}\ \emph {et~al.}(2019)\citenamefont {Yoo},
  \citenamefont {Engelke}, \citenamefont {Carr}, \citenamefont {Fang},
  \citenamefont {Zhang}, \citenamefont {Cazeaux}, \citenamefont {Sung},
  \citenamefont {Hovden}, \citenamefont {Tsen}, \citenamefont {Taniguchi},
  \citenamefont {Watanabe}, \citenamefont {Yi}, \citenamefont {Kim},
  \citenamefont {Luskin}, \citenamefont {Tadmor}, \citenamefont {Kaxiras},\
  and\ \citenamefont {Kim}}]{Yoo2019}%
  \BibitemOpen
  \bibfield  {author} {\bibinfo {author} {\bibfnamefont {H.}~\bibnamefont
  {Yoo}}, \bibinfo {author} {\bibfnamefont {R.}~\bibnamefont {Engelke}},
  \bibinfo {author} {\bibfnamefont {S.}~\bibnamefont {Carr}}, \bibinfo {author}
  {\bibfnamefont {S.}~\bibnamefont {Fang}}, \bibinfo {author} {\bibfnamefont
  {K.}~\bibnamefont {Zhang}}, \bibinfo {author} {\bibfnamefont
  {P.}~\bibnamefont {Cazeaux}}, \bibinfo {author} {\bibfnamefont {S.~H.}\
  \bibnamefont {Sung}}, \bibinfo {author} {\bibfnamefont {R.}~\bibnamefont
  {Hovden}}, \bibinfo {author} {\bibfnamefont {A.~W.}\ \bibnamefont {Tsen}},
  \bibinfo {author} {\bibfnamefont {T.}~\bibnamefont {Taniguchi}}, \bibinfo
  {author} {\bibfnamefont {K.}~\bibnamefont {Watanabe}}, \bibinfo {author}
  {\bibfnamefont {G.-C.}\ \bibnamefont {Yi}}, \bibinfo {author} {\bibfnamefont
  {M.}~\bibnamefont {Kim}}, \bibinfo {author} {\bibfnamefont {M.}~\bibnamefont
  {Luskin}}, \bibinfo {author} {\bibfnamefont {E.~B.}\ \bibnamefont {Tadmor}},
  \bibinfo {author} {\bibfnamefont {E.}~\bibnamefont {Kaxiras}}, \ and\
  \bibinfo {author} {\bibfnamefont {P.}~\bibnamefont {Kim}},\ }\href {\doibase
  10.1038/s41563-019-0346-z} {\bibfield  {journal} {\bibinfo  {journal} {Nat.
  Mater.}\ }\textbf {\bibinfo {volume} {18}},\ \bibinfo {pages} {448} (\bibinfo
  {year} {2019})}\BibitemShut {NoStop}%
\bibitem [{\citenamefont {Walet}\ and\ \citenamefont
  {Guinea}(2019)}]{Walet2019}%
  \BibitemOpen
  \bibfield  {author} {\bibinfo {author} {\bibfnamefont {N.~R.}\ \bibnamefont
  {Walet}}\ and\ \bibinfo {author} {\bibfnamefont {F.}~\bibnamefont {Guinea}},\
  }\href {\doibase 10.1088/2053-1583/ab57f8} {\bibfield  {journal} {\bibinfo
  {journal} {2D Mater.}\ }\textbf {\bibinfo {volume} {7}},\ \bibinfo {pages}
  {015023} (\bibinfo {year} {2019})}\BibitemShut {NoStop}%
\bibitem [{\citenamefont {Alden}\ \emph {et~al.}(2013)\citenamefont {Alden},
  \citenamefont {Tsen}, \citenamefont {Huang}, \citenamefont {Hovden},
  \citenamefont {Brown}, \citenamefont {Park}, \citenamefont {Muller},\ and\
  \citenamefont {McEuen}}]{Alden2013}%
  \BibitemOpen
  \bibfield  {author} {\bibinfo {author} {\bibfnamefont {J.~S.}\ \bibnamefont
  {Alden}}, \bibinfo {author} {\bibfnamefont {A.~W.}\ \bibnamefont {Tsen}},
  \bibinfo {author} {\bibfnamefont {P.~Y.}\ \bibnamefont {Huang}}, \bibinfo
  {author} {\bibfnamefont {R.}~\bibnamefont {Hovden}}, \bibinfo {author}
  {\bibfnamefont {L.}~\bibnamefont {Brown}}, \bibinfo {author} {\bibfnamefont
  {J.}~\bibnamefont {Park}}, \bibinfo {author} {\bibfnamefont {D.~A.}\
  \bibnamefont {Muller}}, \ and\ \bibinfo {author} {\bibfnamefont {P.~L.}\
  \bibnamefont {McEuen}},\ }\href {\doibase 10.1073/pnas.1309394110} {\bibfield
   {journal} {\bibinfo  {journal} {Proc. Natl. Acad. Sci.}\ }\textbf {\bibinfo
  {volume} {110}},\ \bibinfo {pages} {11256} (\bibinfo {year}
  {2013})}\BibitemShut {NoStop}%
\bibitem [{\citenamefont {Martin}\ \emph {et~al.}(2008)\citenamefont {Martin},
  \citenamefont {Blanter},\ and\ \citenamefont {Morpurgo}}]{Martin2008}%
  \BibitemOpen
  \bibfield  {author} {\bibinfo {author} {\bibfnamefont {I.}~\bibnamefont
  {Martin}}, \bibinfo {author} {\bibfnamefont {Y.~M.}\ \bibnamefont {Blanter}},
  \ and\ \bibinfo {author} {\bibfnamefont {A.~F.}\ \bibnamefont {Morpurgo}},\
  }\href {\doibase 10.1103/PhysRevLett.100.036804} {\bibfield  {journal}
  {\bibinfo  {journal} {Phys. Rev. Lett.}\ }\textbf {\bibinfo {volume} {100}},\
  \bibinfo {pages} {036804} (\bibinfo {year} {2008})}\BibitemShut {NoStop}%
\bibitem [{\citenamefont {Zhang}\ \emph {et~al.}(2013)\citenamefont {Zhang},
  \citenamefont {MacDonald},\ and\ \citenamefont {Mele}}]{Zhang2013}%
  \BibitemOpen
  \bibfield  {author} {\bibinfo {author} {\bibfnamefont {F.}~\bibnamefont
  {Zhang}}, \bibinfo {author} {\bibfnamefont {A.~H.}\ \bibnamefont
  {MacDonald}}, \ and\ \bibinfo {author} {\bibfnamefont {E.~J.}\ \bibnamefont
  {Mele}},\ }\href {\doibase 10.1073/pnas.1308853110} {\bibfield  {journal}
  {\bibinfo  {journal} {Proc. Natl. Acad. Sci.}\ }\textbf {\bibinfo {volume}
  {110}},\ \bibinfo {pages} {10546} (\bibinfo {year} {2013})}\BibitemShut
  {NoStop}%
\bibitem [{\citenamefont {Vaezi}\ \emph {et~al.}(2013)\citenamefont {Vaezi},
  \citenamefont {Liang}, \citenamefont {Ngai}, \citenamefont {Yang},\ and\
  \citenamefont {Kim}}]{Vaezi2013}%
  \BibitemOpen
  \bibfield  {author} {\bibinfo {author} {\bibfnamefont {A.}~\bibnamefont
  {Vaezi}}, \bibinfo {author} {\bibfnamefont {Y.}~\bibnamefont {Liang}},
  \bibinfo {author} {\bibfnamefont {D.~H.}\ \bibnamefont {Ngai}}, \bibinfo
  {author} {\bibfnamefont {L.}~\bibnamefont {Yang}}, \ and\ \bibinfo {author}
  {\bibfnamefont {E.-A.}\ \bibnamefont {Kim}},\ }\href {\doibase
  10.1103/PhysRevX.3.021018} {\bibfield  {journal} {\bibinfo  {journal} {Phys.
  Rev. X}\ }\textbf {\bibinfo {volume} {3}},\ \bibinfo {pages} {021018}
  (\bibinfo {year} {2013})}\BibitemShut {NoStop}%
\bibitem [{\citenamefont {Ju}\ \emph {et~al.}(2015)\citenamefont {Ju},
  \citenamefont {Shi}, \citenamefont {Nair}, \citenamefont {Lv}, \citenamefont
  {Jin}, \citenamefont {Velasco}, \citenamefont {Ojeda-Aristizabal},
  \citenamefont {Bechtel}, \citenamefont {Martin}, \citenamefont {Zettl},
  \citenamefont {Analytis},\ and\ \citenamefont {Wang}}]{Ju2015}%
  \BibitemOpen
  \bibfield  {author} {\bibinfo {author} {\bibfnamefont {L.}~\bibnamefont
  {Ju}}, \bibinfo {author} {\bibfnamefont {Z.}~\bibnamefont {Shi}}, \bibinfo
  {author} {\bibfnamefont {N.}~\bibnamefont {Nair}}, \bibinfo {author}
  {\bibfnamefont {Y.}~\bibnamefont {Lv}}, \bibinfo {author} {\bibfnamefont
  {C.}~\bibnamefont {Jin}}, \bibinfo {author} {\bibfnamefont {J.}~\bibnamefont
  {Velasco}}, \bibinfo {author} {\bibfnamefont {C.}~\bibnamefont
  {Ojeda-Aristizabal}}, \bibinfo {author} {\bibfnamefont {H.~A.}\ \bibnamefont
  {Bechtel}}, \bibinfo {author} {\bibfnamefont {M.~C.}\ \bibnamefont {Martin}},
  \bibinfo {author} {\bibfnamefont {A.}~\bibnamefont {Zettl}}, \bibinfo
  {author} {\bibfnamefont {J.}~\bibnamefont {Analytis}}, \ and\ \bibinfo
  {author} {\bibfnamefont {F.}~\bibnamefont {Wang}},\ }\href {\doibase
  10.1038/nature14364} {\bibfield  {journal} {\bibinfo  {journal} {Nature}\
  }\textbf {\bibinfo {volume} {520}},\ \bibinfo {pages} {650} (\bibinfo {year}
  {2015})}\BibitemShut {NoStop}%
\bibitem [{\citenamefont {Yin}\ \emph {et~al.}(2016)\citenamefont {Yin},
  \citenamefont {Jiang}, \citenamefont {Qiao},\ and\ \citenamefont
  {He}}]{Yin2016}%
  \BibitemOpen
  \bibfield  {author} {\bibinfo {author} {\bibfnamefont {L.-J.}\ \bibnamefont
  {Yin}}, \bibinfo {author} {\bibfnamefont {H.}~\bibnamefont {Jiang}}, \bibinfo
  {author} {\bibfnamefont {J.-B.}\ \bibnamefont {Qiao}}, \ and\ \bibinfo
  {author} {\bibfnamefont {L.}~\bibnamefont {He}},\ }\href {\doibase
  10.1038/ncomms11760} {\bibfield  {journal} {\bibinfo  {journal} {Nat.
  Commun.}\ }\textbf {\bibinfo {volume} {7}},\ \bibinfo {pages} {11760}
  (\bibinfo {year} {2016})}\BibitemShut {NoStop}%
\bibitem [{\citenamefont {San-Jose}\ and\ \citenamefont
  {Prada}(2013)}]{San-Jose2013}%
  \BibitemOpen
  \bibfield  {author} {\bibinfo {author} {\bibfnamefont {P.}~\bibnamefont
  {San-Jose}}\ and\ \bibinfo {author} {\bibfnamefont {E.}~\bibnamefont
  {Prada}},\ }\href {\doibase 10.1103/PhysRevB.88.121408} {\bibfield  {journal}
  {\bibinfo  {journal} {Phys. Rev. B}\ }\textbf {\bibinfo {volume} {88}},\
  \bibinfo {pages} {121408(R)} (\bibinfo {year} {2013})}\BibitemShut {NoStop}%
\bibitem [{\citenamefont {Efimkin}\ and\ \citenamefont
  {MacDonald}(2018)}]{Efimkin2018}%
  \BibitemOpen
  \bibfield  {author} {\bibinfo {author} {\bibfnamefont {D.~K.}\ \bibnamefont
  {Efimkin}}\ and\ \bibinfo {author} {\bibfnamefont {A.~H.}\ \bibnamefont
  {MacDonald}},\ }\href {\doibase 10.1103/PhysRevB.98.035404} {\bibfield
  {journal} {\bibinfo  {journal} {Phys. Rev. B}\ }\textbf {\bibinfo {volume}
  {98}},\ \bibinfo {pages} {035404} (\bibinfo {year} {2018})}\BibitemShut
  {NoStop}%
\bibitem [{\citenamefont {Ramires}\ and\ \citenamefont
  {Lado}(2018)}]{Ramires2018}%
  \BibitemOpen
  \bibfield  {author} {\bibinfo {author} {\bibfnamefont {A.}~\bibnamefont
  {Ramires}}\ and\ \bibinfo {author} {\bibfnamefont {J.~L.}\ \bibnamefont
  {Lado}},\ }\href {\doibase 10.1103/PhysRevLett.121.146801} {\bibfield
  {journal} {\bibinfo  {journal} {Phys. Rev. Lett.}\ }\textbf {\bibinfo
  {volume} {121}},\ \bibinfo {pages} {146801} (\bibinfo {year}
  {2018})}\BibitemShut {NoStop}%
\bibitem [{\citenamefont {Huang}\ \emph {et~al.}(2018)\citenamefont {Huang},
  \citenamefont {Kim}, \citenamefont {Efimkin}, \citenamefont {Lovorn},
  \citenamefont {Taniguchi}, \citenamefont {Watanabe}, \citenamefont
  {MacDonald}, \citenamefont {Tutuc},\ and\ \citenamefont {LeRoy}}]{Huang2018}%
  \BibitemOpen
  \bibfield  {author} {\bibinfo {author} {\bibfnamefont {S.}~\bibnamefont
  {Huang}}, \bibinfo {author} {\bibfnamefont {K.}~\bibnamefont {Kim}}, \bibinfo
  {author} {\bibfnamefont {D.~K.}\ \bibnamefont {Efimkin}}, \bibinfo {author}
  {\bibfnamefont {T.}~\bibnamefont {Lovorn}}, \bibinfo {author} {\bibfnamefont
  {T.}~\bibnamefont {Taniguchi}}, \bibinfo {author} {\bibfnamefont
  {K.}~\bibnamefont {Watanabe}}, \bibinfo {author} {\bibfnamefont {A.~H.}\
  \bibnamefont {MacDonald}}, \bibinfo {author} {\bibfnamefont {E.}~\bibnamefont
  {Tutuc}}, \ and\ \bibinfo {author} {\bibfnamefont {B.~J.}\ \bibnamefont
  {LeRoy}},\ }\href {\doibase 10.1103/PhysRevLett.121.037702} {\bibfield
  {journal} {\bibinfo  {journal} {Phys. Rev. Lett.}\ }\textbf {\bibinfo
  {volume} {121}},\ \bibinfo {pages} {037702} (\bibinfo {year}
  {2018})}\BibitemShut {NoStop}%
\bibitem [{\citenamefont {Sunku}\ \emph {et~al.}(2018)\citenamefont {Sunku},
  \citenamefont {Ni}, \citenamefont {Jiang}, \citenamefont {Yoo}, \citenamefont
  {Sternbach}, \citenamefont {McLeod}, \citenamefont {Stauber}, \citenamefont
  {Xiong}, \citenamefont {Taniguchi}, \citenamefont {Watanabe}, \citenamefont
  {Kim}, \citenamefont {Fogler},\ and\ \citenamefont {Basov}}]{Sunku2018}%
  \BibitemOpen
  \bibfield  {author} {\bibinfo {author} {\bibfnamefont {S.~S.}\ \bibnamefont
  {Sunku}}, \bibinfo {author} {\bibfnamefont {G.~X.}\ \bibnamefont {Ni}},
  \bibinfo {author} {\bibfnamefont {B.~Y.}\ \bibnamefont {Jiang}}, \bibinfo
  {author} {\bibfnamefont {H.}~\bibnamefont {Yoo}}, \bibinfo {author}
  {\bibfnamefont {A.}~\bibnamefont {Sternbach}}, \bibinfo {author}
  {\bibfnamefont {A.~S.}\ \bibnamefont {McLeod}}, \bibinfo {author}
  {\bibfnamefont {T.}~\bibnamefont {Stauber}}, \bibinfo {author} {\bibfnamefont
  {L.}~\bibnamefont {Xiong}}, \bibinfo {author} {\bibfnamefont
  {T.}~\bibnamefont {Taniguchi}}, \bibinfo {author} {\bibfnamefont
  {K.}~\bibnamefont {Watanabe}}, \bibinfo {author} {\bibfnamefont
  {P.}~\bibnamefont {Kim}}, \bibinfo {author} {\bibfnamefont {M.~M.}\
  \bibnamefont {Fogler}}, \ and\ \bibinfo {author} {\bibfnamefont {D.~N.}\
  \bibnamefont {Basov}},\ }\href {\doibase 10.1126/science.aau5144} {\bibfield
  {journal} {\bibinfo  {journal} {Science}\ }\textbf {\bibinfo {volume}
  {362}},\ \bibinfo {pages} {1153} (\bibinfo {year} {2018})}\BibitemShut
  {NoStop}%
\bibitem [{\citenamefont {Rickhaus}\ \emph {et~al.}(2018)\citenamefont
  {Rickhaus}, \citenamefont {Wallbank}, \citenamefont {Slizovskiy},
  \citenamefont {Pisoni}, \citenamefont {Overweg}, \citenamefont {Lee},
  \citenamefont {Eich}, \citenamefont {Liu}, \citenamefont {Watanabe},
  \citenamefont {Taniguchi}, \citenamefont {Ihn},\ and\ \citenamefont
  {Ensslin}}]{Rickhaus2018}%
  \BibitemOpen
  \bibfield  {author} {\bibinfo {author} {\bibfnamefont {P.}~\bibnamefont
  {Rickhaus}}, \bibinfo {author} {\bibfnamefont {J.}~\bibnamefont {Wallbank}},
  \bibinfo {author} {\bibfnamefont {S.}~\bibnamefont {Slizovskiy}}, \bibinfo
  {author} {\bibfnamefont {R.}~\bibnamefont {Pisoni}}, \bibinfo {author}
  {\bibfnamefont {H.}~\bibnamefont {Overweg}}, \bibinfo {author} {\bibfnamefont
  {Y.}~\bibnamefont {Lee}}, \bibinfo {author} {\bibfnamefont {M.}~\bibnamefont
  {Eich}}, \bibinfo {author} {\bibfnamefont {M.-H.}\ \bibnamefont {Liu}},
  \bibinfo {author} {\bibfnamefont {K.}~\bibnamefont {Watanabe}}, \bibinfo
  {author} {\bibfnamefont {T.}~\bibnamefont {Taniguchi}}, \bibinfo {author}
  {\bibfnamefont {T.}~\bibnamefont {Ihn}}, \ and\ \bibinfo {author}
  {\bibfnamefont {K.}~\bibnamefont {Ensslin}},\ }\href {\doibase
  10.1021/acs.nanolett.8b02387} {\bibfield  {journal} {\bibinfo  {journal}
  {Nano Lett.}\ }\textbf {\bibinfo {volume} {18}},\ \bibinfo {pages} {6725}
  (\bibinfo {year} {2018})}\BibitemShut {NoStop}%
\bibitem [{\citenamefont {Xu}\ \emph {et~al.}(2019)\citenamefont {Xu},
  \citenamefont {Berdyugin}, \citenamefont {Kumaravadivel}, \citenamefont
  {Guinea}, \citenamefont {{Krishna Kumar}}, \citenamefont {Bandurin},
  \citenamefont {Morozov}, \citenamefont {Kuang}, \citenamefont {Tsim},
  \citenamefont {Liu}, \citenamefont {Edgar}, \citenamefont {Grigorieva},
  \citenamefont {Fal'ko}, \citenamefont {Kim},\ and\ \citenamefont
  {Geim}}]{Xu2019}%
  \BibitemOpen
  \bibfield  {author} {\bibinfo {author} {\bibfnamefont {S.~G.}\ \bibnamefont
  {Xu}}, \bibinfo {author} {\bibfnamefont {A.~I.}\ \bibnamefont {Berdyugin}},
  \bibinfo {author} {\bibfnamefont {P.}~\bibnamefont {Kumaravadivel}}, \bibinfo
  {author} {\bibfnamefont {F.}~\bibnamefont {Guinea}}, \bibinfo {author}
  {\bibfnamefont {R.}~\bibnamefont {{Krishna Kumar}}}, \bibinfo {author}
  {\bibfnamefont {D.~A.}\ \bibnamefont {Bandurin}}, \bibinfo {author}
  {\bibfnamefont {S.~V.}\ \bibnamefont {Morozov}}, \bibinfo {author}
  {\bibfnamefont {W.}~\bibnamefont {Kuang}}, \bibinfo {author} {\bibfnamefont
  {B.}~\bibnamefont {Tsim}}, \bibinfo {author} {\bibfnamefont {S.}~\bibnamefont
  {Liu}}, \bibinfo {author} {\bibfnamefont {J.~H.}\ \bibnamefont {Edgar}},
  \bibinfo {author} {\bibfnamefont {I.~V.}\ \bibnamefont {Grigorieva}},
  \bibinfo {author} {\bibfnamefont {V.~I.}\ \bibnamefont {Fal'ko}}, \bibinfo
  {author} {\bibfnamefont {M.}~\bibnamefont {Kim}}, \ and\ \bibinfo {author}
  {\bibfnamefont {A.~K.}\ \bibnamefont {Geim}},\ }\href {\doibase
  10.1038/s41467-019-11971-7} {\bibfield  {journal} {\bibinfo  {journal} {Nat.
  Commun.}\ }\textbf {\bibinfo {volume} {10}},\ \bibinfo {pages} {4008}
  (\bibinfo {year} {2019})}\BibitemShut {NoStop}%
\bibitem [{\citenamefont {Chalker}\ and\ \citenamefont
  {Coddington}(1988)}]{Chalker1988}%
  \BibitemOpen
  \bibfield  {author} {\bibinfo {author} {\bibfnamefont {J.~T.}\ \bibnamefont
  {Chalker}}\ and\ \bibinfo {author} {\bibfnamefont {P.~D.}\ \bibnamefont
  {Coddington}},\ }\href {\doibase 10.1088/0022-3719/21/14/008} {\bibfield
  {journal} {\bibinfo  {journal} {J. Phys. C Solid State Phys.}\ }\textbf
  {\bibinfo {volume} {21}},\ \bibinfo {pages} {2665} (\bibinfo {year}
  {1988})}\BibitemShut {NoStop}%
\bibitem [{\citenamefont {Fleischmann}\ \emph {et~al.}(2020)\citenamefont
  {Fleischmann}, \citenamefont {Gupta}, \citenamefont {Wullschl{\"{a}}ger},
  \citenamefont {Theil}, \citenamefont {Weckbecker}, \citenamefont {Meded},
  \citenamefont {Sharma}, \citenamefont {Meyer},\ and\ \citenamefont
  {Shallcross}}]{Fleischmann2020}%
  \BibitemOpen
  \bibfield  {author} {\bibinfo {author} {\bibfnamefont {M.}~\bibnamefont
  {Fleischmann}}, \bibinfo {author} {\bibfnamefont {R.}~\bibnamefont {Gupta}},
  \bibinfo {author} {\bibfnamefont {F.}~\bibnamefont {Wullschl{\"{a}}ger}},
  \bibinfo {author} {\bibfnamefont {S.}~\bibnamefont {Theil}}, \bibinfo
  {author} {\bibfnamefont {D.}~\bibnamefont {Weckbecker}}, \bibinfo {author}
  {\bibfnamefont {V.}~\bibnamefont {Meded}}, \bibinfo {author} {\bibfnamefont
  {S.}~\bibnamefont {Sharma}}, \bibinfo {author} {\bibfnamefont
  {B.}~\bibnamefont {Meyer}}, \ and\ \bibinfo {author} {\bibfnamefont
  {S.}~\bibnamefont {Shallcross}},\ }\href {\doibase
  10.1021/acs.nanolett.9b04027} {\bibfield  {journal} {\bibinfo  {journal}
  {Nano Lett.}\ }\textbf {\bibinfo {volume} {20}},\ \bibinfo {pages} {971}
  (\bibinfo {year} {2020})}\BibitemShut {NoStop}%
\bibitem [{\citenamefont {Tsim}\ \emph {et~al.}(2020)\citenamefont {Tsim},
  \citenamefont {Nam},\ and\ \citenamefont {Koshino}}]{Tsim2020}%
  \BibitemOpen
  \bibfield  {author} {\bibinfo {author} {\bibfnamefont {B.}~\bibnamefont
  {Tsim}}, \bibinfo {author} {\bibfnamefont {N.~N.~T.}\ \bibnamefont {Nam}}, \
  and\ \bibinfo {author} {\bibfnamefont {M.}~\bibnamefont {Koshino}},\ }\href
  {\doibase 10.1103/PhysRevB.101.125409} {\bibfield  {journal} {\bibinfo
  {journal} {Phys. Rev. B}\ }\textbf {\bibinfo {volume} {101}},\ \bibinfo
  {pages} {125409} (\bibinfo {year} {2020})}\BibitemShut {NoStop}%
\bibitem [{\citenamefont {{De Beule}}\ \emph {et~al.}(2020)\citenamefont {{De
  Beule}}, \citenamefont {Dominguez},\ and\ \citenamefont
  {Recher}}]{DeBeule2020}%
  \BibitemOpen
  \bibfield  {author} {\bibinfo {author} {\bibfnamefont {C.}~\bibnamefont {{De
  Beule}}}, \bibinfo {author} {\bibfnamefont {F.}~\bibnamefont {Dominguez}}, \
  and\ \bibinfo {author} {\bibfnamefont {P.}~\bibnamefont {Recher}},\ }\href
  {\doibase 10.1103/PhysRevLett.125.096402} {\bibfield  {journal} {\bibinfo
  {journal} {Phys. Rev. Lett.}\ }\textbf {\bibinfo {volume} {125}},\ \bibinfo
  {pages} {096402} (\bibinfo {year} {2020})}\BibitemShut {NoStop}%
\bibitem [{\citenamefont {Kramer}\ \emph {et~al.}(2005)\citenamefont {Kramer},
  \citenamefont {Ohtsuki},\ and\ \citenamefont {Kettemann}}]{Kramer2005}%
  \BibitemOpen
  \bibfield  {author} {\bibinfo {author} {\bibfnamefont {B.}~\bibnamefont
  {Kramer}}, \bibinfo {author} {\bibfnamefont {T.}~\bibnamefont {Ohtsuki}}, \
  and\ \bibinfo {author} {\bibfnamefont {S.}~\bibnamefont {Kettemann}},\ }\href
  {\doibase 10.1016/j.physrep.2005.07.001} {\bibfield  {journal} {\bibinfo
  {journal} {Phys. Rep.}\ }\textbf {\bibinfo {volume} {417}},\ \bibinfo {pages}
  {211} (\bibinfo {year} {2005})}\BibitemShut {NoStop}%
\bibitem [{\citenamefont {Mkhitaryan}\ and\ \citenamefont
  {Raikh}(2009)}]{Mkhitaryan2009}%
  \BibitemOpen
  \bibfield  {author} {\bibinfo {author} {\bibfnamefont {V.~V.}\ \bibnamefont
  {Mkhitaryan}}\ and\ \bibinfo {author} {\bibfnamefont {M.~E.}\ \bibnamefont
  {Raikh}},\ }\href {\doibase 10.1103/PhysRevB.79.125401} {\bibfield  {journal}
  {\bibinfo  {journal} {Phys. Rev. B}\ }\textbf {\bibinfo {volume} {79}},\
  \bibinfo {pages} {125401} (\bibinfo {year} {2009})}\BibitemShut {NoStop}%
\bibitem [{\citenamefont {Lee}(1994)}]{Lee1994}%
  \BibitemOpen
  \bibfield  {author} {\bibinfo {author} {\bibfnamefont {D.~H.}\ \bibnamefont
  {Lee}},\ }\href {\doibase 10.1103/PhysRevB.50.10788} {\bibfield  {journal}
  {\bibinfo  {journal} {Phys. Rev. B}\ }\textbf {\bibinfo {volume} {50}},\
  \bibinfo {pages} {10788} (\bibinfo {year} {1994})}\BibitemShut {NoStop}%
\bibitem [{\citenamefont {Pasek}\ and\ \citenamefont
  {Chong}(2014)}]{Pasek2014}%
  \BibitemOpen
  \bibfield  {author} {\bibinfo {author} {\bibfnamefont {M.}~\bibnamefont
  {Pasek}}\ and\ \bibinfo {author} {\bibfnamefont {Y.~D.}\ \bibnamefont
  {Chong}},\ }\href {https://link.aps.org/doi/10.1103/PhysRevB.89.075113}
  {\bibfield  {journal} {\bibinfo  {journal} {Phys. Rev. B}\ }\textbf {\bibinfo
  {volume} {89}},\ \bibinfo {pages} {075113} (\bibinfo {year}
  {2014})}\BibitemShut {NoStop}%
\bibitem [{\citenamefont {Delplace}\ \emph {et~al.}(2017)\citenamefont
  {Delplace}, \citenamefont {Fruchart},\ and\ \citenamefont
  {Tauber}}]{Delplace2017}%
  \BibitemOpen
  \bibfield  {author} {\bibinfo {author} {\bibfnamefont {P.}~\bibnamefont
  {Delplace}}, \bibinfo {author} {\bibfnamefont {M.}~\bibnamefont {Fruchart}},
  \ and\ \bibinfo {author} {\bibfnamefont {C.}~\bibnamefont {Tauber}},\ }\href
  {\doibase 10.1103/PhysRevB.95.205413} {\bibfield  {journal} {\bibinfo
  {journal} {Phys. Rev. B}\ }\textbf {\bibinfo {volume} {95}},\ \bibinfo
  {pages} {205413} (\bibinfo {year} {2017})}\BibitemShut {NoStop}%
\bibitem [{\citenamefont {Delplace}(2020)}]{Delplace2020}%
  \BibitemOpen
  \bibfield  {author} {\bibinfo {author} {\bibfnamefont {P.}~\bibnamefont
  {Delplace}},\ }\href {\doibase 10.21468/SciPostPhys.8.5.081} {\bibfield
  {journal} {\bibinfo  {journal} {SciPost Phys.}\ }\textbf {\bibinfo {volume}
  {8}},\ \bibinfo {pages} {081} (\bibinfo {year} {2020})}\BibitemShut {NoStop}%
\bibitem [{\citenamefont {Kitagawa}\ \emph {et~al.}(2010)\citenamefont
  {Kitagawa}, \citenamefont {Berg}, \citenamefont {Rudner},\ and\ \citenamefont
  {Demler}}]{Kitagawa2010}%
  \BibitemOpen
  \bibfield  {author} {\bibinfo {author} {\bibfnamefont {T.}~\bibnamefont
  {Kitagawa}}, \bibinfo {author} {\bibfnamefont {E.}~\bibnamefont {Berg}},
  \bibinfo {author} {\bibfnamefont {M.}~\bibnamefont {Rudner}}, \ and\ \bibinfo
  {author} {\bibfnamefont {E.}~\bibnamefont {Demler}},\ }\href {\doibase
  10.1103/PhysRevB.82.235114} {\bibfield  {journal} {\bibinfo  {journal} {Phys.
  Rev. B}\ }\textbf {\bibinfo {volume} {82}},\ \bibinfo {pages} {235114}
  (\bibinfo {year} {2010})}\BibitemShut {NoStop}%
\bibitem [{\citenamefont {Liang}\ and\ \citenamefont
  {Chong}(2013)}]{Liang2013}%
  \BibitemOpen
  \bibfield  {author} {\bibinfo {author} {\bibfnamefont {G.~Q.}\ \bibnamefont
  {Liang}}\ and\ \bibinfo {author} {\bibfnamefont {Y.~D.}\ \bibnamefont
  {Chong}},\ }\href {\doibase 10.1103/PhysRevLett.110.203904} {\bibfield
  {journal} {\bibinfo  {journal} {Phys. Rev. Lett.}\ }\textbf {\bibinfo
  {volume} {110}},\ \bibinfo {pages} {203904} (\bibinfo {year}
  {2013})}\BibitemShut {NoStop}%
\bibitem [{\citenamefont {Rudner}\ \emph {et~al.}(2013)\citenamefont {Rudner},
  \citenamefont {Lindner}, \citenamefont {Berg},\ and\ \citenamefont
  {Levin}}]{Rudner2014}%
  \BibitemOpen
  \bibfield  {author} {\bibinfo {author} {\bibfnamefont {M.~S.}\ \bibnamefont
  {Rudner}}, \bibinfo {author} {\bibfnamefont {N.~H.}\ \bibnamefont {Lindner}},
  \bibinfo {author} {\bibfnamefont {E.}~\bibnamefont {Berg}}, \ and\ \bibinfo
  {author} {\bibfnamefont {M.}~\bibnamefont {Levin}},\ }\href {\doibase
  10.1103/PhysRevX.3.031005} {\bibfield  {journal} {\bibinfo  {journal} {Phys.
  Rev. X}\ }\textbf {\bibinfo {volume} {3}},\ \bibinfo {pages} {031005}
  (\bibinfo {year} {2013})}\BibitemShut {NoStop}%
\bibitem [{\citenamefont {Mukherjee}\ and\ \citenamefont
  {Rechtsman}(2020)}]{Mukherjee2020}%
  \BibitemOpen
  \bibfield  {author} {\bibinfo {author} {\bibfnamefont {S.}~\bibnamefont
  {Mukherjee}}\ and\ \bibinfo {author} {\bibfnamefont {M.~C.}\ \bibnamefont
  {Rechtsman}},\ }\href {\doibase 10.1126/science.aba8725} {\bibfield
  {journal} {\bibinfo  {journal} {Science}\ }\textbf {\bibinfo {volume}
  {368}},\ \bibinfo {pages} {856} (\bibinfo {year} {2020})}\BibitemShut
  {NoStop}%
\bibitem [{\citenamefont {Chou}\ \emph {et~al.}(2020)\citenamefont {Chou},
  \citenamefont {Wu},\ and\ \citenamefont {{Das Sarma}}}]{Chou2020}%
  \BibitemOpen
  \bibfield  {author} {\bibinfo {author} {\bibfnamefont {Y.-Z.}\ \bibnamefont
  {Chou}}, \bibinfo {author} {\bibfnamefont {F.}~\bibnamefont {Wu}}, \ and\
  \bibinfo {author} {\bibfnamefont {S.}~\bibnamefont {{Das Sarma}}},\ }\href
  {\doibase 10.1103/PhysRevResearch.2.033271} {\bibfield  {journal} {\bibinfo
  {journal} {Phys. Rev. Res.}\ }\textbf {\bibinfo {volume} {2}},\ \bibinfo
  {pages} {033271} (\bibinfo {year} {2020})}\BibitemShut {NoStop}%
\bibitem [{\citenamefont {Potter}\ \emph {et~al.}(2020)\citenamefont {Potter},
  \citenamefont {Chalker},\ and\ \citenamefont {Gurarie}}]{Potter2020}%
  \BibitemOpen
  \bibfield  {author} {\bibinfo {author} {\bibfnamefont {A.~C.}\ \bibnamefont
  {Potter}}, \bibinfo {author} {\bibfnamefont {J.~T.}\ \bibnamefont {Chalker}},
  \ and\ \bibinfo {author} {\bibfnamefont {V.}~\bibnamefont {Gurarie}},\ }\href
  {\doibase 10.1103/PhysRevLett.125.086601} {\bibfield  {journal} {\bibinfo
  {journal} {Phys. Rev. Lett.}\ }\textbf {\bibinfo {volume} {125}},\ \bibinfo
  {pages} {086601} (\bibinfo {year} {2020})}\BibitemShut {NoStop}%
\bibitem [{\citenamefont {Raghu}\ and\ \citenamefont
  {Haldane}(2008)}]{Raghu2008}%
  \BibitemOpen
  \bibfield  {author} {\bibinfo {author} {\bibfnamefont {S.}~\bibnamefont
  {Raghu}}\ and\ \bibinfo {author} {\bibfnamefont {F.~D.~M.}\ \bibnamefont
  {Haldane}},\ }\href {\doibase 10.1103/PhysRevA.78.033834} {\bibfield
  {journal} {\bibinfo  {journal} {Phys. Rev. A}\ }\textbf {\bibinfo {volume}
  {78}},\ \bibinfo {pages} {033834} (\bibinfo {year} {2008})}\BibitemShut
  {NoStop}%
\bibitem [{\citenamefont {Lu}\ \emph {et~al.}(2014)\citenamefont {Lu},
  \citenamefont {Joannopoulos},\ and\ \citenamefont
  {Solja{\v{c}}i{\'{c}}}}]{Lu2014}%
  \BibitemOpen
  \bibfield  {author} {\bibinfo {author} {\bibfnamefont {L.}~\bibnamefont
  {Lu}}, \bibinfo {author} {\bibfnamefont {J.~D.}\ \bibnamefont
  {Joannopoulos}}, \ and\ \bibinfo {author} {\bibfnamefont {M.}~\bibnamefont
  {Solja{\v{c}}i{\'{c}}}},\ }\href {\doibase 10.1038/nphoton.2014.248}
  {\bibfield  {journal} {\bibinfo  {journal} {Nat. Photonics}\ }\textbf
  {\bibinfo {volume} {8}},\ \bibinfo {pages} {821} (\bibinfo {year}
  {2014})}\BibitemShut {NoStop}%
\bibitem [{\citenamefont {Ozawa}\ \emph {et~al.}(2019)\citenamefont {Ozawa},
  \citenamefont {Price}, \citenamefont {Amo}, \citenamefont {Goldman},
  \citenamefont {Hafezi}, \citenamefont {Lu}, \citenamefont {Rechtsman},
  \citenamefont {Schuster}, \citenamefont {Simon}, \citenamefont {Zilberberg},\
  and\ \citenamefont {Carusotto}}]{Ozawa2019}%
  \BibitemOpen
  \bibfield  {author} {\bibinfo {author} {\bibfnamefont {T.}~\bibnamefont
  {Ozawa}}, \bibinfo {author} {\bibfnamefont {H.~M.}\ \bibnamefont {Price}},
  \bibinfo {author} {\bibfnamefont {A.}~\bibnamefont {Amo}}, \bibinfo {author}
  {\bibfnamefont {N.}~\bibnamefont {Goldman}}, \bibinfo {author} {\bibfnamefont
  {M.}~\bibnamefont {Hafezi}}, \bibinfo {author} {\bibfnamefont
  {L.}~\bibnamefont {Lu}}, \bibinfo {author} {\bibfnamefont {M.~C.}\
  \bibnamefont {Rechtsman}}, \bibinfo {author} {\bibfnamefont {D.}~\bibnamefont
  {Schuster}}, \bibinfo {author} {\bibfnamefont {J.}~\bibnamefont {Simon}},
  \bibinfo {author} {\bibfnamefont {O.}~\bibnamefont {Zilberberg}}, \ and\
  \bibinfo {author} {\bibfnamefont {I.}~\bibnamefont {Carusotto}},\ }\href
  {\doibase 10.1103/RevModPhys.91.015006} {\bibfield  {journal} {\bibinfo
  {journal} {Rev. Mod. Phys.}\ }\textbf {\bibinfo {volume} {91}},\ \bibinfo
  {pages} {015006} (\bibinfo {year} {2019})}\BibitemShut {NoStop}%
\bibitem [{\citenamefont {Zhong}\ \emph {et~al.}(2020)\citenamefont {Zhong},
  \citenamefont {Li}, \citenamefont {Song}, \citenamefont {Kartashov},
  \citenamefont {Zhang}, \citenamefont {Zhang},\ and\ \citenamefont
  {Chen}}]{Zhong2020}%
  \BibitemOpen
  \bibfield  {author} {\bibinfo {author} {\bibfnamefont {H.}~\bibnamefont
  {Zhong}}, \bibinfo {author} {\bibfnamefont {Y.}~\bibnamefont {Li}}, \bibinfo
  {author} {\bibfnamefont {D.}~\bibnamefont {Song}}, \bibinfo {author}
  {\bibfnamefont {Y.~V.}\ \bibnamefont {Kartashov}}, \bibinfo {author}
  {\bibfnamefont {Y.}~\bibnamefont {Zhang}}, \bibinfo {author} {\bibfnamefont
  {Y.}~\bibnamefont {Zhang}}, \ and\ \bibinfo {author} {\bibfnamefont
  {Z.}~\bibnamefont {Chen}},\ }\href {\doibase 10.1002/lpor.202000001}
  {\bibfield  {journal} {\bibinfo  {journal} {Laser \& Photonics Reviews}\
  }\textbf {\bibinfo {volume} {14}},\ \bibinfo {pages} {2000001} (\bibinfo
  {year} {2020})}\BibitemShut {NoStop}%
\bibitem [{\citenamefont {Lewenstein}\ \emph {et~al.}(2007)\citenamefont
  {Lewenstein}, \citenamefont {Sanpera}, \citenamefont {Ahufinger},
  \citenamefont {Damski}, \citenamefont {Sen(De)},\ and\ \citenamefont
  {Sen}}]{Lewenstein2007}%
  \BibitemOpen
  \bibfield  {author} {\bibinfo {author} {\bibfnamefont {M.}~\bibnamefont
  {Lewenstein}}, \bibinfo {author} {\bibfnamefont {A.}~\bibnamefont {Sanpera}},
  \bibinfo {author} {\bibfnamefont {V.}~\bibnamefont {Ahufinger}}, \bibinfo
  {author} {\bibfnamefont {B.}~\bibnamefont {Damski}}, \bibinfo {author}
  {\bibfnamefont {A.}~\bibnamefont {Sen(De)}}, \ and\ \bibinfo {author}
  {\bibfnamefont {U.}~\bibnamefont {Sen}},\ }\href {\doibase
  10.1080/00018730701223200} {\bibfield  {journal} {\bibinfo  {journal}
  {Advances in Physics}\ }\textbf {\bibinfo {volume} {56}},\ \bibinfo {pages}
  {243} (\bibinfo {year} {2007})}\BibitemShut {NoStop}%
\bibitem [{\citenamefont {Bloch}\ \emph {et~al.}(2008)\citenamefont {Bloch},
  \citenamefont {Dalibard},\ and\ \citenamefont {Zwerger}}]{Bloch2008}%
  \BibitemOpen
  \bibfield  {author} {\bibinfo {author} {\bibfnamefont {I.}~\bibnamefont
  {Bloch}}, \bibinfo {author} {\bibfnamefont {J.}~\bibnamefont {Dalibard}}, \
  and\ \bibinfo {author} {\bibfnamefont {W.}~\bibnamefont {Zwerger}},\ }\href
  {\doibase 10.1103/RevModPhys.80.885} {\bibfield  {journal} {\bibinfo
  {journal} {Rev. Mod. Phys.}\ }\textbf {\bibinfo {volume} {80}},\ \bibinfo
  {pages} {885} (\bibinfo {year} {2008})}\BibitemShut {NoStop}%
\bibitem [{\citenamefont {Bakr}\ \emph {et~al.}(2009)\citenamefont {Bakr},
  \citenamefont {Gillen}, \citenamefont {Peng}, \citenamefont {F{\"o}lling},\
  and\ \citenamefont {Greiner}}]{Bakr2009}%
  \BibitemOpen
  \bibfield  {author} {\bibinfo {author} {\bibfnamefont {W.~S.}\ \bibnamefont
  {Bakr}}, \bibinfo {author} {\bibfnamefont {J.~I.}\ \bibnamefont {Gillen}},
  \bibinfo {author} {\bibfnamefont {A.}~\bibnamefont {Peng}}, \bibinfo {author}
  {\bibfnamefont {S.}~\bibnamefont {F{\"o}lling}}, \ and\ \bibinfo {author}
  {\bibfnamefont {M.}~\bibnamefont {Greiner}},\ }\href {\doibase
  10.1038/nature08482} {\bibfield  {journal} {\bibinfo  {journal} {Nature}\
  }\textbf {\bibinfo {volume} {462}},\ \bibinfo {pages} {74} (\bibinfo {year}
  {2009})}\BibitemShut {NoStop}%
\bibitem [{\citenamefont {Wintersperger}\ \emph {et~al.}(2020)\citenamefont
  {Wintersperger}, \citenamefont {Braun}, \citenamefont {{\"{U}}nal},
  \citenamefont {Eckardt}, \citenamefont {Liberto}, \citenamefont {Goldman},
  \citenamefont {Bloch},\ and\ \citenamefont
  {Aidelsburger}}]{Wintersperger2020}%
  \BibitemOpen
  \bibfield  {author} {\bibinfo {author} {\bibfnamefont {K.}~\bibnamefont
  {Wintersperger}}, \bibinfo {author} {\bibfnamefont {C.}~\bibnamefont
  {Braun}}, \bibinfo {author} {\bibfnamefont {F.~N.}\ \bibnamefont
  {{\"{U}}nal}}, \bibinfo {author} {\bibfnamefont {A.}~\bibnamefont {Eckardt}},
  \bibinfo {author} {\bibfnamefont {M.~D.}\ \bibnamefont {Liberto}}, \bibinfo
  {author} {\bibfnamefont {N.}~\bibnamefont {Goldman}}, \bibinfo {author}
  {\bibfnamefont {I.}~\bibnamefont {Bloch}}, \ and\ \bibinfo {author}
  {\bibfnamefont {M.}~\bibnamefont {Aidelsburger}},\ }\href {\doibase
  10.1038/s41567-020-0949-y} {\bibfield  {journal} {\bibinfo  {journal} {Nat.
  Phys.}\ }\textbf {\bibinfo {volume} {16}},\ \bibinfo {pages} {1058} (\bibinfo
  {year} {2020})}\BibitemShut {NoStop}%
\bibitem [{\citenamefont {Zou}\ \emph {et~al.}(2018)\citenamefont {Zou},
  \citenamefont {Po}, \citenamefont {Vishwanath},\ and\ \citenamefont
  {Senthil}}]{Zou2018}%
  \BibitemOpen
  \bibfield  {author} {\bibinfo {author} {\bibfnamefont {L.}~\bibnamefont
  {Zou}}, \bibinfo {author} {\bibfnamefont {H.~C.}\ \bibnamefont {Po}},
  \bibinfo {author} {\bibfnamefont {A.}~\bibnamefont {Vishwanath}}, \ and\
  \bibinfo {author} {\bibfnamefont {T.}~\bibnamefont {Senthil}},\ }\href
  {\doibase 10.1103/PhysRevB.98.085435} {\bibfield  {journal} {\bibinfo
  {journal} {Phys. Rev. B}\ }\textbf {\bibinfo {volume} {98}},\ \bibinfo
  {pages} {085435} (\bibinfo {year} {2018})}\BibitemShut {NoStop}%
\bibitem [{\citenamefont {Qiao}\ \emph {et~al.}(2014)\citenamefont {Qiao},
  \citenamefont {Jung}, \citenamefont {Lin}, \citenamefont {Ren}, \citenamefont
  {MacDonald},\ and\ \citenamefont {Niu}}]{Qiao2014a}%
  \BibitemOpen
  \bibfield  {author} {\bibinfo {author} {\bibfnamefont {Z.}~\bibnamefont
  {Qiao}}, \bibinfo {author} {\bibfnamefont {J.}~\bibnamefont {Jung}}, \bibinfo
  {author} {\bibfnamefont {C.}~\bibnamefont {Lin}}, \bibinfo {author}
  {\bibfnamefont {Y.}~\bibnamefont {Ren}}, \bibinfo {author} {\bibfnamefont
  {A.~H.}\ \bibnamefont {MacDonald}}, \ and\ \bibinfo {author} {\bibfnamefont
  {Q.}~\bibnamefont {Niu}},\ }\href {\doibase 10.1103/PhysRevLett.112.206601}
  {\bibfield  {journal} {\bibinfo  {journal} {Phys. Rev. Lett.}\ }\textbf
  {\bibinfo {volume} {112}},\ \bibinfo {pages} {206601} (\bibinfo {year}
  {2014})}\BibitemShut {NoStop}%
\bibitem [{\citenamefont {L\'opez~Sancho}\ \emph {et~al.}(1985)\citenamefont
  {L\'opez~Sancho}, \citenamefont {L\'opez~Sancho},\ and\ \citenamefont
  {Rubio}}]{LopezSancho1985}%
  \BibitemOpen
  \bibfield  {author} {\bibinfo {author} {\bibfnamefont {M.~P.}\ \bibnamefont
  {L\'opez~Sancho}}, \bibinfo {author} {\bibfnamefont {J.~M.}\ \bibnamefont
  {L\'opez~Sancho}}, \ and\ \bibinfo {author} {\bibfnamefont {J.}~\bibnamefont
  {Rubio}},\ }\href {\doibase 10.1088/0305-4608/15/4/009} {\bibfield  {journal}
  {\bibinfo  {journal} {J. Phys. F Met. Phys.}\ }\textbf {\bibinfo {volume}
  {15}},\ \bibinfo {pages} {851} (\bibinfo {year} {1985})}\BibitemShut
  {NoStop}%
\bibitem [{\citenamefont {Wu}\ and\ \citenamefont {Cao}(2008)}]{Wu2008}%
  \BibitemOpen
  \bibfield  {author} {\bibinfo {author} {\bibfnamefont {B.~H.}\ \bibnamefont
  {Wu}}\ and\ \bibinfo {author} {\bibfnamefont {J.~C.}\ \bibnamefont {Cao}},\
  }\href {https://doi.org/10.1088/0953-8984/20/8/085224} {\bibfield  {journal}
  {\bibinfo  {journal} {J. Phys. Condens. Matter}\ }\textbf {\bibinfo {volume}
  {20}},\ \bibinfo {pages} {085224} (\bibinfo {year} {2008})}\BibitemShut
  {NoStop}%
\bibitem [{\citenamefont {Moskalets}\ and\ \citenamefont
  {B{\"{u}}ttiker}(2002)}]{Moskalets2002}%
  \BibitemOpen
  \bibfield  {author} {\bibinfo {author} {\bibfnamefont {M.}~\bibnamefont
  {Moskalets}}\ and\ \bibinfo {author} {\bibfnamefont {M.}~\bibnamefont
  {B{\"{u}}ttiker}},\ }\href {\doibase 10.1103/PhysRevB.66.205320} {\bibfield
  {journal} {\bibinfo  {journal} {Phys. Rev. B}\ }\textbf {\bibinfo {volume}
  {66}},\ \bibinfo {pages} {205320} (\bibinfo {year} {2002})}\BibitemShut
  {NoStop}%
\bibitem [{\citenamefont {Camalet}\ \emph {et~al.}(2003)\citenamefont
  {Camalet}, \citenamefont {Lehmann}, \citenamefont {Kohler},\ and\
  \citenamefont {H{\"{a}}nggi}}]{Camalet2003}%
  \BibitemOpen
  \bibfield  {author} {\bibinfo {author} {\bibfnamefont {S.}~\bibnamefont
  {Camalet}}, \bibinfo {author} {\bibfnamefont {J.}~\bibnamefont {Lehmann}},
  \bibinfo {author} {\bibfnamefont {S.}~\bibnamefont {Kohler}}, \ and\ \bibinfo
  {author} {\bibfnamefont {P.}~\bibnamefont {H{\"{a}}nggi}},\ }\href {\doibase
  10.1103/PhysRevLett.90.210602} {\bibfield  {journal} {\bibinfo  {journal}
  {Phys. Rev. Lett.}\ }\textbf {\bibinfo {volume} {90}},\ \bibinfo {pages}
  {210602} (\bibinfo {year} {2003})}\BibitemShut {NoStop}%
\bibitem [{\citenamefont {{Foa Torres}}\ \emph {et~al.}(2014)\citenamefont
  {{Foa Torres}}, \citenamefont {{Perez-Piskunow}}, \citenamefont {Balseiro},\
  and\ \citenamefont {Usaj}}]{FoaTorres2014}%
  \BibitemOpen
  \bibfield  {author} {\bibinfo {author} {\bibfnamefont {L.~E.~F.}\
  \bibnamefont {{Foa Torres}}}, \bibinfo {author} {\bibfnamefont {P.~M.}\
  \bibnamefont {{Perez-Piskunow}}}, \bibinfo {author} {\bibfnamefont {C.~A.}\
  \bibnamefont {Balseiro}}, \ and\ \bibinfo {author} {\bibfnamefont
  {G.}~\bibnamefont {Usaj}},\ }\href {\doibase 10.1103/PhysRevLett.113.266801}
  {\bibfield  {journal} {\bibinfo  {journal} {Phys. Rev. Lett.}\ }\textbf
  {\bibinfo {volume} {113}},\ \bibinfo {pages} {266801} (\bibinfo {year}
  {2014})}\BibitemShut {NoStop}%
\bibitem [{\citenamefont {Farrell}\ and\ \citenamefont
  {Pereg-Barnea}(2016)}]{Farrell2016}%
  \BibitemOpen
  \bibfield  {author} {\bibinfo {author} {\bibfnamefont {A.}~\bibnamefont
  {Farrell}}\ and\ \bibinfo {author} {\bibfnamefont {T.}~\bibnamefont
  {Pereg-Barnea}},\ }\href {\doibase 10.1103/PhysRevB.93.045121} {\bibfield
  {journal} {\bibinfo  {journal} {Phys. Rev. B}\ }\textbf {\bibinfo {volume}
  {93}},\ \bibinfo {pages} {045121} (\bibinfo {year} {2016})}\BibitemShut
  {NoStop}%
\end{thebibliography}%

\end{document}